\documentclass{article} 
\usepackage{lineno,hyperref}
\usepackage{amssymb,amsthm}
\usepackage{amsfonts}
\modulolinenumbers[5]
 
\DeclareMathAlphabet{\itbf}{OML}{cmm}{b}{it}
     
 \usepackage{authblk}     
     
\usepackage{graphicx,color}     
\newcommand{\re}{ \textcolor{black}}

    \title{Emergence of Turbulent Epochs in  Oil Prices} 
      
 \author[1,3]{Josselin Garnier}
\author[2,3]{Knut Solna}
 
\affil[1]{\footnotesize Centre de Math\'ematiques Appliqu\'ees, Ecole Polytechnique, 91128 Palaiseau Cedex, France}
\affil[2]{\footnotesize    Department of Mathematics,  University of California, Irvine CA 92697}
  \affil[3]{\footnotesize  Lusenn, 3 Place de l'Eglise,  29570 Roscanvel, France}    
      
\begin{document}
 
\maketitle 
 
\begin{abstract}
Oil price data have a complicated multi-scale structure that may vary with  time. 
We use time-frequency analysis to identify the main features of these variations and, in particular, the regime shifts. 
The analysis is based on a wavelet-based decomposition and analysis of the associated scale spectrum.
The joint estimation of the local Hurst exponent and volatility is the key to detect and identify regime shifting and switching  of the oil price.
The framework involves in particular modeling 
in terms of a process of ``multi-fractional''  type so that both the roughness and the volatility of the price process
may vary with time. Special epochs then emerge 
as a result of these degrees of freedom, moreover, 
as a  result of the special type of spectral estimator used. 
These special epochs are discussed and related to historical events.
Some of them are not detected by standard analysis based on maximum likelihood estimation.
The paper  presents a novel algorithm for robust detection of such special epochs and multi-fractional behavior in 
financial or other types of data.  
In the financial context  insight about such behavior of the asset price is important to evaluate financial contracts involving the asset.
\end{abstract}


 \section{Introduction}  
 
 It   has been well known since the pioneering work of Mandelbrot \cite{mandelbrot67,mandelbrot71}
that market price fluctuations are  poorly modeled and described by discrete time random walks or continuous diffusions
driven by standard Brownian motions.
We  refer to \cite{mandelbrot97} for a historical presentation.
In this paper we consider oil price data for the period  from May 1987 to June 2016. The prices are recorded every day.
We seek to understand the time-frequency character of the data.
As we show below, the spectral characteristics of the data may change over time and, in particular, signify regime shifts. 
The identification of transitions in the commodity market structure is indeed of utmost importance for hedging and risk assessment \cite{engle14}.

In this paper we describe the scale spectrum of the oil price data, 
which is the main tool we use to unravel the time-frequency structures of the data \cite{b2,b14}.
The scale spectrum of a (locally) stationary process is essentially its (local) power spectral density
\re{as a function of scale}, which are the reciprocals of frequencies.  \re{
The scale spectrum analysis shows that the oil price data exhibit a power-law character, in that its scale spectrum 
obeys a power law.  This is similar to the type of scaling one sees, for instance, in turbulence data \cite{b19,PS}.   
Power law type  scaling behavior has been observed in many other physical 
systems also, see for instance \cite{wind,temp}.  
See also  \cite{gao1} for a comprehensive overview of such
modeling and discussions of various applications and estimation approaches.}
What is somewhat surprising with the oil price data is that this power law persists over many scales, in fact, essentially over all the available scales. 
The power-law parameters we discuss here are the Hurst exponent and volatility.
The Hurst exponent determines how price changes over different time intervals are correlated, 
and it also characterizes the power law of the scale spectrum as a \re{function of scale}. 
The volatility determines the typical magnitude of the relative price changes.  
The power-law character of the scale spectrum, however, varies over time, and, as mentioned, suggests regime shifts. 
The variations in the power-law parameters reveal periods in the data that cannot be seen directly.
It is also of interest to look at scale-based cross spectra which give information
about possible time-varying collective behavior in the multivariate case \cite{cross} and we will do so below.

The estimation method for the power-law parameters  of oil price data 
that we propose is based on a wavelet decomposition.
It has already been proposed in the literature to estimate the local fractal dimension, or Hurst exponent, of 
different sets of data, either synthetic time series \cite{b10,moulines,b7} or experimental (physical or financial) time series  
\cite{audit02,bacry03,bacry10,bayraktar,b23b,simonsen}, and to use wavelet-based decomposition to do so. 
It has long been identified that wavelet analysis is an important addition to time-series methods with practical applications
in economics and finance \cite{ramsey02,gallegati}. 
For instance wavelets were used to study the evolution of the impact of oil price changes in
the macroeconomy \cite{aguiar11}, 
 to estimate the Hurst exponent of the crude oil price \cite{elder08} and 
to investigate the issue of market efficiency in futures markets for oil \cite{yousefi05}.
 \re{  Market efficiency is commonly related to the degree that the price returns
 have  a random character.  Indeed the Hurst exponent has been used to
quantify this since it is a measure of correlation
or coherence in between returns, and therefore their 
``predictability''  based on historical returns.  
Other measures such as entropy-based measures  
have also been used as measures for predictability.  
In \cite{gao3} a permutation entropy measure was used to identify 
special epochs in the Chinese stock market by observing this measure associated 
with some main indices as functions of time.  In \cite{gao3}  the authors consider
minute scale data and compute the permutation entropy measure for each day.
This measure is indeed lower during the epochs of some main market events.   
Another entropy-based measure is presented in \cite{yang1,yang2,yang4,yang5}. 
Here the measure is scale-based in that it corresponds
to an entropy measure with respect to the series of partial data sums of different lengths.
The scaling of the entropy can then be related to the power-law parameters.
A point being made in these papers is that the nonlinear  nature of the entropy measure
calls for a bias correction which is identified and used.     
Various data sets are used, ranging from financial index data to physiological data and user online behavior,
and in these various data sets interesting power law behaviors are identified,
which supports the importance and relevance of such modeling. Further issues of bias correction
for power-law type estimation are presented in \cite{yang3,yang6} 
where it is shown that this step may be important when dealing with relatively short data.   
Here we choose a wavelet-based  Hurst exponent estimation method to relate the modeling to  a correlation measure
and have a flexible multiscale approach that allows for theoretical
characterization of performance for an ``ideal''   power-law like fractional 
Brownian motion, as well as cross-wavelet analysis.    
Other approaches  to estimate the Hurst exponent are discussed in  for instance \cite{alvarez02,cajueiro04,serletis04,jiang14}.  
See also \cite{gao2} for a comparison of methods and discussion of another interesting application where the break in
 the scaling structure comes from a target embedded in sea clutter radar data.  } 

\re{The wavelet-based method for the joint estimation of the Hurst exponent and the volatility that we propose is 
original and the analysis of the two parameters reveals a more detailed structure.
From the estimation point of view, the joint estimation method that we propose is original 
in that we use continuous or non-decimated wavelet coefficients that are strongly correlated.
From the practical point of view, 
the joint analysis of the two parameters allows for sharper detection and identification of regime switches.
An important aspect of the estimator we set forth is its robustness which makes it 
superior in the context of real data  as compared to the optimal scheme associated with noise 
free or ideal data, we discuss this in \ref{sec:appA}.  
 There, we discuss in particular how our algorithm performs better in the case of noisy data 
 than an estimator designed to be optimal in the noise free case. 
In \cite{lahmiri,l2} the presence  of chaos and non-linear dynamics, 
in particular in the crude oil market,  is examined with a view
toward identifying  change of dynamics connected  to financial crises.
The robust estimator and modeling we use here allow us to identify  fine grained 
events of financial and economic importance associated with the crude oil price over the last 
 30 years period with a view toward the specific correlation
structure that is important in financial applications. 
Some of the special periods that are revealed by our analysis can be related to historical events
(such as the Asian financial crisis in July 1997 or the Russian financial crisis in August 1998).
It is worth noticing that a local volatility estimation or a local Hurst exponent estimation by maximum likelihood do not detect these events,
as we will see below.
Our analysis also reveals a special period in 2000 which does not seem to be related to any major historical event.}

We remark that when modeling with time series in general the presence, or not, of
 a power-law  correlation structure is crucially important. The case of  temporal variations
 in the volatility, in particular stochastic volatility models in finance, 
 have been considered in a huge number of papers in last decades, see for instance
 \cite{booksv} for an overview. What  we find from the
 analysis in this paper is a corroboration of the notion that  modeling with
 a Hurst exponent that may change,  corresponding to different roughness of the process
 in different epochs, is important \cite{bacry10,asym}. 
 We remark that a number of recent studies have considered such behavior 
 in the context of the bitcoin market \cite{bit1,bit2}. 
  Indeed, for risk calculations and 
 pricing problems in a  
 financial context the value of the Hurst exponent and memory scaling
 is very  important \cite{price}. 
 This is particularly the case when one considers  long time horizons.    
 
The outline of the paper  is as follows. 
In Section~\ref{sec:price}, we plot the scale spectra of the data sets and show that price modeling with a local power-law 
structure is indeed appropriate.
We describe the structure of the Hurst exponent and volatility estimated over successive overlapping windows
in Section~\ref{sec:model}.  
We underline that joint estimation of the two parameters is critical, in particular
for obtaining a correct assessment of the volatility variations.
For comparison, we also study the estimated volatility for the standard (local geometric 
Brownian motion) model in Section~\ref{sec:uncorrelated},
when the Hurst exponent $H$ is assumed to be equal to $1/2$, corresponding to Brownian scaling 
with independent increments or returns.
Moreover, we study natural gas price in Section \ref{sec:gas}, which shows that the procedure is general and
efficient to detect collective or individual events by looking at cross spectra.
Finally, in \ref{app:mBm} till \ref{sec:appC},   we provide  the details of the
modeling, 
the estimation and further figures for respectively the oil and the gas data.

\section{Oil Price Data}
\begin{figure}[htbp]
   \centering
   \includegraphics[width=6.4cm]{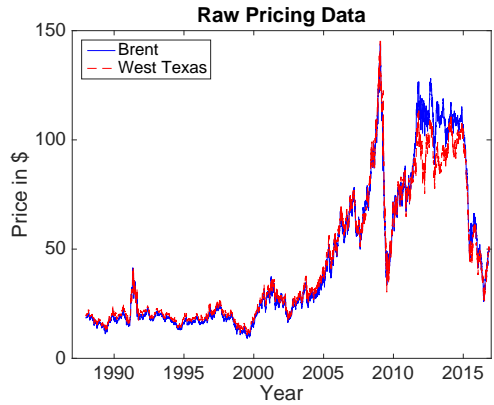} 
   \caption{Full price data vectors for West Texas (dashed red line) and Brent (solid blue line).}
   \label{fig:1}
\end{figure}

\begin{figure}
   \centering
   \includegraphics[width=6.0cm]{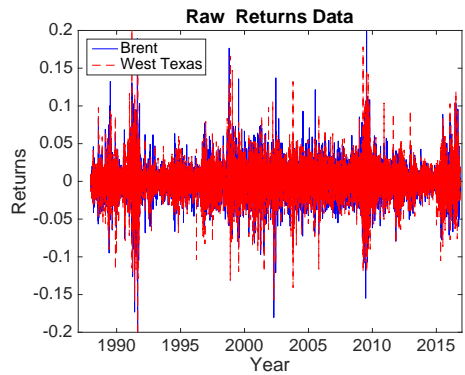} 
   \caption{Returns for West Texas (dashed red line) and Brent (solid blue line).}
   \label{fig:2}
\end{figure}

In Figure \ref{fig:1} the dashed red line shows the raw daily oil price data  for the
 ``Western Texas Intermediate (WTI), Spot Price Free On Board (FOB) (New York)''  (hereafter ``West Texas")
 in dollars per barrel.   
 The solid blue line represents the ``Europe Brent Spot Price FOB (London)"
 (hereafter ``Brent")  in dollars per barrel.
 The daily data are available from May 1997 to June 2016 \cite{data_oil}.
 In Figure  \ref{fig:2}  in  \ref{sec:appB}
 we plot the corresponding returns which shows that the variability in the price changes vary with time. 
 
   \re{
 In Figure \ref{fig:2}, we show the returns process for the Brent and West Texas data sets defined by
 \begin{equation}
 \label{eq:ret}
   R_n  = \frac{ P(t_n+\delta t) - P(t_n) }{P(t_n)} ,  
\end{equation}
where $P(t)$ is the raw price data and
 $t_n=n \delta t$, with  $\delta t$  being one day. 
We observe here from the magnitude of the fluctuations seen in Figure \ref{fig:2} 
that the  volatility  of the returns process is not constant, but rather exhibits temporal variations.
The returns for West Texas and Brent are strongly correlated. In fact this is the case
for the increments at different scales and this is illustrated  in  Figure \ref{fig:corr},
  \ref{sec:appB}.}
   
\section{The Scale Spectrum for Log-Oil Price Data}
\label{sec:price}

\begin{figure}[htbp]
   \centering
      \includegraphics[width=6.4cm]{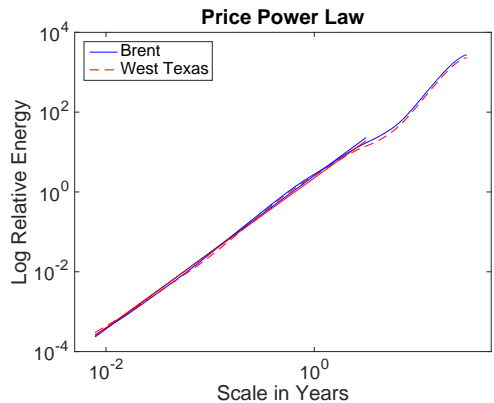} 
   \caption{\re{The  ``global power law'' for West Texas data (dashed red line) and 
      Brent data (solid blue line).  When the scale spectrum is computed from the complete log-transformed data, we observe
      approximately a power law. The estimated Hurst exponents are $H=.46$ (Brent) and $H=.44$ (West Texas). 
          The estimated volatilities are $\sigma=34 \%$ (Brent) and $\sigma= 32 \%$ (West Texas).
           The straight line segments 
       on top represent the  fitted spectra  and fit extremely well the data from the shortest scale to an outer scale
      of about one year. 
      Here and below the spectrum is computed with Haar wavelets and is ``continuous'' in time and scale.
       Note that the value $H=1/2$  corresponds  to 
      the classic case of uncorrelated returns. 
      As discussed in \ref{sec:appA} the precision of the Hurst estimate 
      can be assessed by the precision obtained with  fractional Brownian 
      motion.  For the window size of the global data correpsonding
      to almost $2^{13}$ data points and when using the scale range up to
      approximately 1.5 years (the straight line segments  in the figure) this gives
      a relative precision of about 4\% and a negligible bias for the estimate of the Hurst
      coefficient. 
       Thus,    the estimates values  for  $H$ is significantly below $H=1/2$. }
    }
   \label{fig:3}
\end{figure}

 We show in Figure  \ref{fig:3}   the scale  spectra for the log-transformed data  for West Texas (dashed red line) and 
Brent (solid blue line).
For each scale the spectral value is  computed as the local average of 
squares of the wavelet coefficients corresponding to that scale.
 We use the ``Haar'' wavelet basis, making it robust with respect to noise,
as we show in the  \ref{sec:appA}.
 The first level Haar coefficients correspond to the consecutive differences  in the data.
In our case, the data is the log prices so that the Haar  coefficients form an analogue
of the returns process. The Haar coefficients at higher levels correspond  to differences in local 
averages of increasing length. Thus, the higher order differences may be thought of as returns 
over longer intervals.     

In Figure~\ref{fig:3}, we can see a linear behavior in the log-log plot corresponding
 to a power law, making it possible to discuss the power-law parameters, which are the Hurst exponent $H$
 and volatility $\sigma$.
 The Hurst exponent characterizes the power-law decay of the scale spectrum.
 It also characterizes the existence of correlations between the increments of the process.
 If $H=1/2$, then the increments are not correlated, which is the typical case for Brownian motion 
 or a similar diffusion process.
 If $H>1/2$, the increments are positively correlated, which is the phenomenon called ``persistence".
 If $H<1/2$, then the increments are  negatively correlated, which is the phenomenon called ``anti-persistence".
 The  smoothness of the process increases with $H$, as consecutive increments become
 more correlated as $H$ increases. 
    
In  Figure~\ref{fig:3}  
the scale spectra conform with a global power law,
      with estimated Hurst exponents $H=.46$ (Brent) and $H=.44$ (West Texas),
      and estimated volatilities $\sigma=34 \%$ (Brent) and $\sigma= 32 \%$ (West Texas).
      As we will see below, this global power law is in fact 
      consistent with a situation in which the Hurst exponent and volatility vary in a consistent way over subwindows. 
      It is striking to see that a non-trivial power law (that is, a power law with $H \neq 1/2$) 
      emerges very clearly from the financial data, while it is difficult
      to exhibit such a structure from physical data (such as the distribution of energy among turbulence vortices), 
      for which there are, on the contrary, theoretical arguments to support a power law (for instance,  Kolmogorov's theory of turbulence) \cite{PS}.
           
In Figure \ref{fig:4}  in \ref{sec:appB} we plot the scale spectra for the first and last halves of the data. 
They show  that the qualitative behavior of the spectrum  is similar for  the two  halves of the data,
but the parameters vary over time.

 \section{Price Modeling with a Local Power-Law Spectral Structure}
 \label{sec:model}

  The log of the oil price data $P(t)$
  exhibits a global power-law spectral structure, but the more detailed analysis that we report here reveals that the parameters of  the
  power law vary over time. 
  We therefore model the log prices as:
  \begin{equation}
  \label{eq:1}
  \log( P(t)/P(0)  ) =  B_{H,\sigma}(t )\, ,
  \end{equation}
  where $B_{H,\sigma}$ is a random process with local power-law behavior,
  with $\sigma$ being the volatility and $H$ the Hurst exponent. 
  The classic model process for $B_{H,\sigma}$ is  
  fractional Brownian motion (with constant $H$ and $\sigma$) \cite{b18,b29}.
   The parameters
  $\sigma, H$ will be modeled themselves as varying with respect to time.
  This type of modeling is referred to as multi-fractal or multi-fractional stochastic modeling
  \cite{benassi97,vehel95} and is reviewed in  \ref{app:mBm}.
   
  In order to identify the local power-law parameters $H$ and $\sigma$, we decompose the data into overlapping windows
  of length $2^8$ points (windows of roughly one year) in the next section and estimate 
  a homogeneous power law within each window.
  We explain in more detail in the  \ref{sec:appA}
  how the scale spectrum and the local power-law 
  parameters are estimated from the price data.
   The estimated power-law parameters
  are then attributed to the date corresponding to the center of the window.

\subsection{Local Hurst Exponents}
\label{sec:H}%
   
\begin{figure}[htbp]
   \centering
   \includegraphics[width=6.4cm]{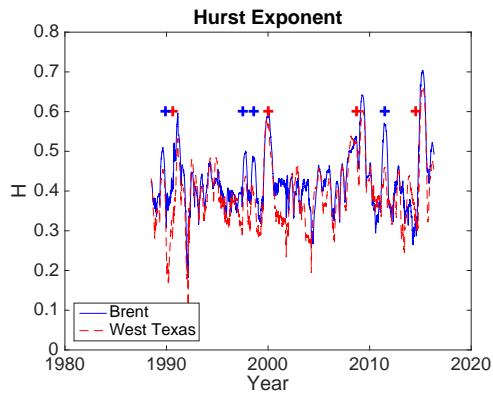}     
    \caption{Estimated Hurst exponents $H_t$
  for the West Texas data (dashed red line)  and  the Brent data  (solid blue line).}
   \label{fig:6}
\end{figure}

In Figure \ref{fig:6}, we show the estimated Hurst exponents.
Recall that the Hurst exponent determines how the consecutive increments of the process
are correlated with values larger than $1/2$,  corresponding to  positive correlation;
values less than $1/2$, corresponding to negative correlation; and $1/2$, corresponding to 
uncorrelated returns and absence of arbitrage. 
In particular, if $H>1/2$, then the market is not weakly efficient as it possesses long memory~\cite{tabak07}.

We use here again the log of the raw pricing data as shown in Figure \ref{fig:1}
when we compute the scale spectrum with the
West Texas data set (dashed red line)  and the Brent data set (solid blue line).
We use  windows of length that represent approximately one year  ($2^{8}$ points).
We move the center point of the window by one day at a time to get a  daily time-series of  the local Hurst exponent. 

\re{Note that with a finite window size the estimates of the parameters will be associated
with sample fluctuations and also a small bias and this should be kept in mind
in the context of Figures \ref{fig:6}-\ref{fig:7a} below. 
One approach for a rough characterization of the sample fluctuations and bias is to compare with the 
standard deviation and bias for this window size when the data is pure  fractional Brownian motion,
an ``ideal'' power law. 
This approach to the characterization of precision is discussed 
in depth in \ref{sec:appA}. The discussion in particular  quantifies 
the sample uncertainty for the Hurst exponent estimate  with the window size
considered here ($2^{8}$ points) to be about 12\%. 
We stress here that the estimator for the power-law parameters uses a special weight matrix 
in the regression of the scale spectrum, see \ref{sec:appA}. As we show 
there this makes the accuracy  comparable to the (optimal) maximum likelihood 
estimator as introduced in \cite{brouste} in the case of noise free data. However, 
the  method we use here gives an estimator that is
both efficient and also robust with respect to model uncertainty, 
contrarily to other estimators, 
this aspect is also discussed in \ref{sec:appA}.
  }
 
Note that in the figure, the four primary periods with high  Hurst exponent estimates are 
 roughly 1990--1991, 1999--2000, 2008--2009, and 2014--2015 and they are indicated by red
 crosses. 
These are important events
exhibited by this time-frequency analysis. We comment more
on this below, given that the four periods are even more visible when looking at the volatilities.  
Some of these periods are neither apparent in the raw price data directly 
(Figure \ref{fig:1}), 
nor in the standard volatilities estimated with standard quadratic variations
(which means assuming that the Hurst exponent is $1/2$), as we will see below.
We also remark that the Hurst exponent is slightly
 lower in the  West Texas data set than in the Brent data set, corresponding to the price
 fluctuations being somewhat  rougher in the West Texas data  than in the Brent data.   
\re{Such a situation with important market events being reflected by the (local) power law 
behavior is consistent with  results also from other data sets, see for instance
\cite{gao3,yang1}. } 

\subsection{Local Volatilities}
\label{sec:ret}%

\begin{figure}[htbp]
   \centering
   \includegraphics[width=6.4cm]{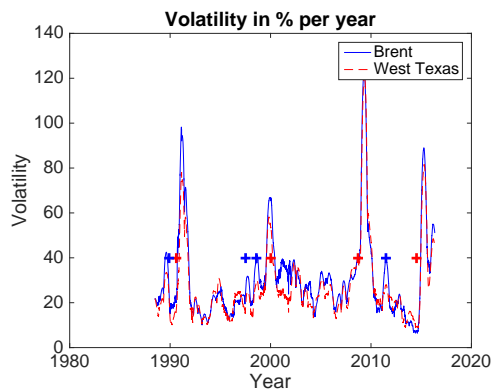}  
      \caption{Local  volatility estimates relative to  the annual time scale  $\sigma_t$
   for the West Texas data (dashed red line)
   and the Brent data (solid blue line).}
   \label{fig:7a}
\end{figure}

When we analyse the log-transformed oil price data, we simultaneously estimate two 
parameters: the Hurst exponent $H_t$  and the local volatility
parameter $\sigma_t$ in Equation~(\ref{eq:1}). 
Using the same segmentation as the one used to generate Figure \ref{fig:6},  we  show the 
corresponding volatility estimate $\sigma_t$ in Figure \ref{fig:7a}. 
The volatility is given relative to the {annual} time scale.
Note again that the power law should 
be interpreted as a local power law with a volatility that depends on time.   

 As noted above, the figure clearly shows that there are four 
 primary periods, roughly 1990--1991, 1999--2000, 2008--2009, and 2014--2015, with high volatility. 
 These four periods can be related to four events, marked with the red crosses in Figures \ref{fig:6}  and \ref{fig:7a}.\\
 -  The first red cross, in August 1990, corresponds to Iraq's invasion of Kuwait, and it initiates
 a period with high volatility and high Hurst exponent.\\
 -  The second red cross, in January 2000, corresponds to the peak of a period with high volatility and
 high Hurst exponent. We hazard a guess, albeit a speculative one, that this may be explained by the approach of the year 2000 and fear of the Y2K bug that never occurred.\\
- The third red cross, in September 2008, corresponds to the bankruptcy of Lehman Brothers, initiating a period 
 with very high volatility and high Hurst exponent. We can also note that the all-time high for oil price was reached during trading on 11 July 2008.\\
- The fourth red cross, in July 2014,  corresponds to the massive liquidation of Brent- and WTI-linked derivatives by fund managers
 and the beginning of the price fall, initiating a period with very high Hurst exponent and high volatility.\\
 Note that the second period, around 2000, cannot be detected from the direct inspection of
 the raw price data, nor from recent analysis such as the 
  Dynamic Conditional Beta method applied to the Brent data to exhibit shocks to commodity markets \cite{engle14}. 
Furthermore, the fourth (and last) special period appears to be special  as its Hurst exponent reaches $.7$,
 a manifestation of strong, positive correlations between the increments. The latter has never been reported in any financial data
 as far as we know.
  
The results are very much the same for the Brent and West Texas data.
In fact, the detailed analysis of the scale-based correlation structure between the two data sets carried
 out in the  \ref{sec:appB}
 reveals that the data sets are indeed strongly correlated on  all scales.
 
In Figures \ref{fig:6} and \ref{fig:7a} we have also marked with blue crosses
 four secondary events that can be identified in the parameter
 processes.
      \\
 - The first blue  cross, in November 1989,  
 corresponds to the fall  of the Berlin wall.    \\
 - The second  blue  cross, in  July  1997, corresponds to  Asian financial  crisis. \\
 - The third  blue  cross, in  August  1998, corresponds to  the Russian financial crisis.  \\
 - The fourth  blue  cross, in September 2010,  corresponds to the European debt crisis. \\
We can notice that the Brent data show a higher sensitivity to these events
than the West Texas data. This may be related to the fact that Brent reflects the international crude oil demand
and is more sensitive to international events than West Texas that reflects crude oil demand in USA.

The qualitative properties with respect to Hurst exponent and volatility estimates, as well as the special periods, are stable
with respect to segmentation, in that they can also be identified as doubling or halving the window lengths.    
Halving the window length makes the estimates become slightly more noisy, while doubling it causes some of the features to become slightly more blurred, in particular in the case of the 2014--2015 period. 
\re{
 We stress here also  that the list of events discussed above
 that can be detected by our method is not the standard list of events detected by inspection of high volatility or other criteria. In particular, we detect a clear event around 2000 that is not noticeable otherwise.}

\subsection{Spectral Misfits}
\begin{figure}[htbp]
   \centering
   \includegraphics[width=6.4cm]{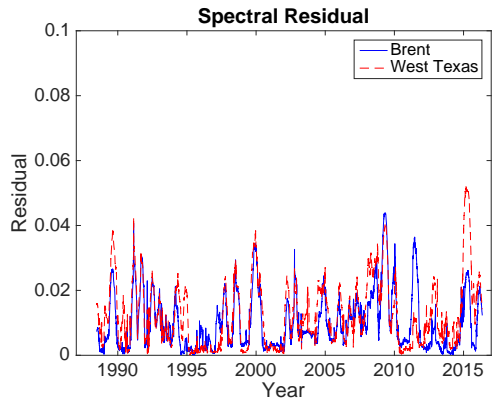} 
   \caption{Spectral misfits for the West Texas data (dashed red line)
   and the Brent data (solid blue line).   }
   \label{fig:7e}
\end{figure}

 Next, we calculate a spectral misfit. This is the root mean square of the differences between
the empirical log-scale spectra and the estimated log-scale spectra (i.e. the power spectra with the estimated
power-law parameters). 
  The result  is plotted in Figure \ref{fig:7e}.
We observe  that  the spectral misfits appear to be  statistically homogeneous with respect to time. 
This means that the four special periods that were detected and discussed above are well-described
by the multi-fractional  model with Hurst exponent $H_t$ and volatility $\sigma_t$.

\section{Modeling with Uncorrelated Returns} 
\label{sec:uncorrelated}

\begin{figure}[htbp]
   \centering
   \includegraphics[width=6.4cm]{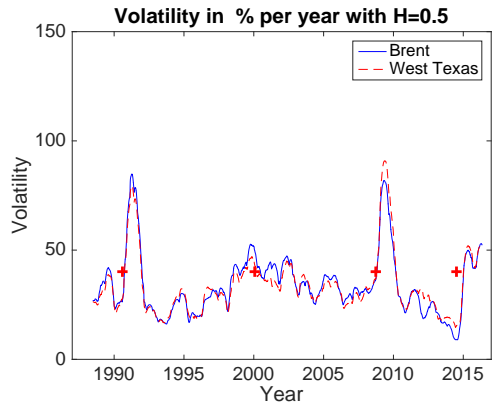}  
   \caption{Estimated volatilities for the West Texas data (dashed red line)
   and the Brent data (solid blue line)  when we condition on $H=1/2$ to enforce uncorrelated returns.   }
   \label{fig:8a}
\end{figure}

In Figure \ref{fig:8a}, we show the estimated volatility when we condition the Hurst exponent $H$ to
be $1/2$, corresponding to Brownian scaling with independent increments, or returns,
which is the standard model.
We can observe that the 1999--2000 period does not appear clearly in this figure, while it does in
Figures \ref{fig:6}-\ref{fig:7a}. 
The 2014--2015 period appears much less dramatic,
while the multi-fractional  analysis reveals its unique features characterized by a very large Hurst exponent. Moreover, the secondary events
marked by blue crosses in Figure \ref{fig:6} are no longer visible, while they are remarkable events.  
Note also that beyond the special periods, the standard volatility experiences somewhat strong variations,
while it is rather constant (around $20\%$)  in the multi-fractional  analysis, except during the special periods.

\begin{figure}[htbp]
   \centering
   \includegraphics[width=6.4cm]{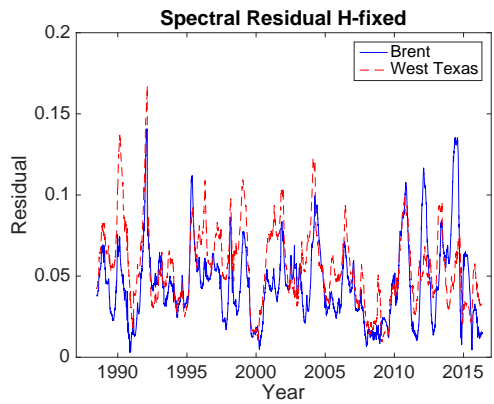} 
   \caption{Spectral misfits for the West Texas data (dashed red line)
   and the Brent data (solid blue line) when we condition on $H=1/2$ to enforce uncorrelated returns.   }
   \label{fig:8c}
\end{figure}

\begin{figure}[htbp]
   \centering
   \includegraphics[width=6.4cm]{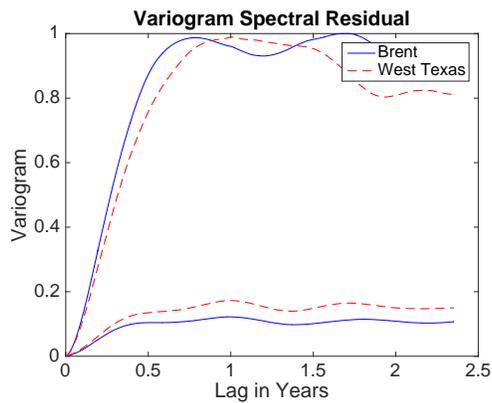} 
   \caption{Variograms for the spectral  misfits. The two top lines correspond to
   the spectral misfits obtained when enforcing uncorrelated returns. 
   The two bottom lines correspond to the spectral misfits obtained  with the multi-fractional model.  }
      \label{fig:8d}
\end{figure}

In Figure \ref{fig:8c}, we show the spectral misfits that follow when we fix $H=1/2$.
Comparing with Figure \ref{fig:7e}, we see that this enforcement means that we do a 
poor job of capturing important structural features in the data, as the spectral misfits are high,
and they can vary significantly during the special periods detected and discussed above.
This is even clearer seen from  the variograms for the spectral misfits   in Figure \ref{fig:8c} (see for instance \cite{vario} for a discussion of the variogram).  
The variograms plotted in Figure \ref{fig:8d}  show  that the spectral misfits in the uncorrelated case are high and have a strong coherence.
 The magnitude of the spectral misfits obtained with the multi-fractional  model
(with a time-varying Hurst exponent) is significantly smaller, and we can see that the spectral misfit
appears to be a white noise process supporting this modeling.

\section{The Scale Spectrum for Natural Gas Price}
\label{sec:gas}%
\begin{figure}[htbp]
   \centering
   \includegraphics[width=6.4cm]{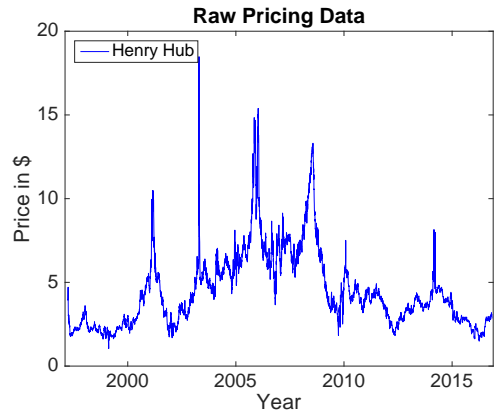} 
   \caption{Henry Hub natural gas spot price.}
   \label{fig:gasprice}
\end{figure}

\begin{figure}[htbp]
   \centering
   \includegraphics[width=6.4cm]{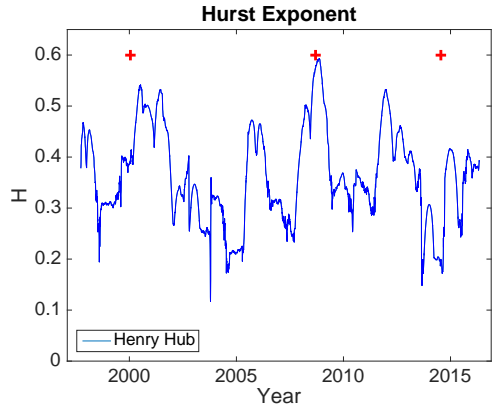} 
   \caption{Estimated Hurst exponents $H_t$
  for the Henry Hub natural gas spot price. The locations of the crosses are as in Figure \ref{fig:6}.}
   \label{fig:9}
\end{figure}

\begin{figure}[htbp]
   \centering
   \includegraphics[width=6.4cm]{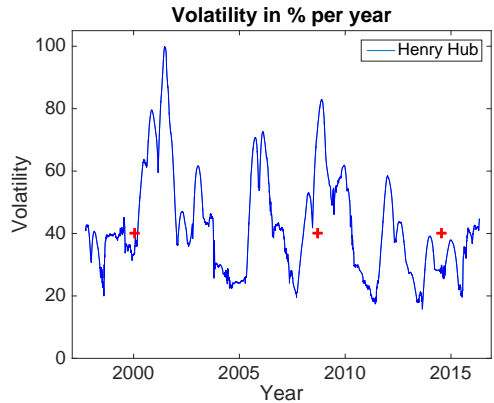} 
   \caption{Estimated volatility $\sigma_t$
  for the Henry Hub natural gas spot price.}
   \label{fig:10}
\end{figure}

In this section we address another set of data that is connected to the previous ones:
the Henry Hub natural gas spot price (Dollars per Million Btu), which is widely used as a U.S. benchmark price for natural gas.
The daily data are available from January 1997 to June 2016 \cite{data_gas} and they are plotted in Figure \ref{fig:gasprice}.
The  price exhibits obvious spikes: the California Energy Crisis
in 2001, a short  price spike  during the week of February 24, 2003 in response to physical market conditions (cold winter) 
leading to low supply and high demand, a price increase at the end of 2005 due to hurricanes Katrina and Rita  
and volatile weather, another price increase in 2008 corresponding to high oil price.
We carry out an analysis similar to the one  for the crude oil price and give the detailed results in the  \ref{sec:appC}.  
We focus in this section on the main results.
  \begin{figure}[htbp]
   \centering
   \includegraphics[width=7.8cm]{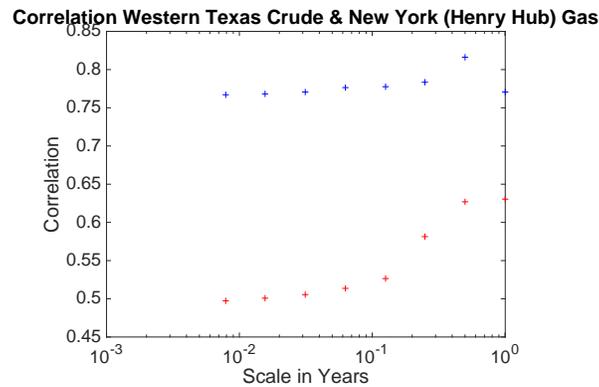} 
   \caption{Scale-based correlation between the West Texas (crude) data and Henry Hub (gas)   data
   during the period January 1997--June 2009 (blue crosses) and July 2009-June 2016 (red crosses). 
   }
   \label{fig:11}
\end{figure}
The time series econometric relationship between the Henry Hub natural gas price 
and the West Texas crude oil price
has been the subject of investigation for a long time  \cite{villar06}.
An  important feature is that the historically strong correlation between oil 
and natural gas prices has recently ceased in North America, 
as natural gas prices have been kept down
in the last period shown 
 by the rapid development of shale gas.
This decoupling is reported by the U.S. Energy Information Administration 
and it has been estimated to happen around 2009 \cite{erdos12}.
In Figure \ref{fig:11} we plot the scale-based correlation  between the West Texas data and Henry Hub data
before and after June 2009. It is clear that the correlation has been dramatically reduced. All scales are affected,
although the scales below one month are more affected.
More strikingly, the time-dependent Hurst exponent and  volatility of the Henry Hub natural gas price exhibit 
the same special periods as for the oil
(characterized by a Hurst exponent larger than $1/2$), 
except the last period 2014--2015. In this period, which corresponds to the highest values 
for the Hurst  exponent of the West Texas data,
the Henry Hub data show no special behavior (see Figures \ref{fig:9}-\ref{fig:10} with
red cross locations as in Figure \ref{fig:6}). 
This is a clear manifestation of the decoupling between oil price and natural gas price.
Finally, note that the short price spikes such as the one during the week of February 24, 2003 do not appear 
at all in the estimated power-law parameters. 
These price spikes are short, they are due to physical market conditions leading to low supply and/or high demand for a 
short time, and they affect the structure of the market price only very locally.



\section{Conclusions and Perspectives}

We have analysed oil price data with a view toward identifying regime switches.
We have found  that  a scale spectral analysis of the log prices is an efficient
approach for identifying regime shifts. The time-frequency analysis involves computing
the scale spectrum  and fitting it to a power law, which for the log-scale spectrum  
corresponds to a linear model (Figure \ref{fig:3}).  
The special regimes can be associated with enhancement of both 
volatility and persistence (Hurst exponent) (Figures \ref{fig:6}-\ref{fig:7a}).

\re{ A striking  result is that the power-law behavior  for the log prices can be seen 
over all scales   available in the data, see Figure \ref{fig:3} where the scales range 
from a couple of days till  about thirty years.  }
 It is very interesting, however, that if one looks at subwindows of the data,
one observes a ``local'' power law with a local Hurst exponent  and  local volatility that vary over time
and, in particular, that  announce  the emergence of  the special regimes
(Figures \ref{fig:6}-\ref{fig:7a}). They fluctuate, however, 
in a coordinated fashion, and  in a way so as to  generate a power law also 
when the data are analyzed globally.
 We refer to the time series defined by the estimates of the Hurst exponent and volatility
in the subwindows as the inferred volatility and Hurst exponent processes.
Note that to generate the inferred processes,  the lengths of the subwindows are fixed,  and we move the window one day at a time.
The choice of the length  of the subwindows is guided by the effective signal-to-noise ratio:
we need to choose windows (a) large enough to have enough data to be able to estimate
the local power-law parameters with sufficient precision, and (b) small enough to resolve  
the local power-law parameters without too much bias.

The results presented here have important consequences in mathematical finance 
modeling and analysis.
 For instance, it is  central to understand and better quantify the arbitrage involved with the 
inferred parameters of the type  observed here. Will a small amount of friction in terms
of a typical transaction cost \cite{schach}, for example, remove the possibility of arbitrage, or is intrinsic arbitrage 
a central ingredient of special regimes as observed here? Moreover, 
can we  
understand  the local power-law behavior we have seen for the price processes discussed here
from  an economic perspective?  
 Finally, we remark that we here looked at daily oil prices with a view toward identifying
 medium and long term structural features in the data.  Additional
 high frequency, or intraday,  price information  give complementary information and indeed is important for traders in these markets.

 \section*{Acknowledgements}
This work was supported by   in part by 
Centre Cournot, Fondation Cournot,  
Universit\'e Paris Saclay (chaire D'Alembert).
A preliminary version of this paper has been published by Centre Cournot \cite{prisme}.
 
\bibliography{proc_oil.bib}
 
\newpage

\appendix

\section{ Multifractional Brownian motion}
\label{app:mBm}
Here we give a precise definition of a multifractional Brownian motion.
It was introduced in \cite{benassi97,vehel95} and more details can be found in \cite{cohenistas} for instance.
Let $H : \mathbb{R} \to (0, 1)$ and $\sigma : \mathbb{R} \to (0,\infty)$ be two measurable functions.
A real-valued process $B_{H,\sigma}(t)$ 
is called a multifractional Brownian motion with Hurst exponent $H$ and volatility $\sigma$
 if it admits the harmonizable representation
\begin{equation}
\label{def:mbm}
 B_{H,\sigma}(t)
 =
\frac{\sigma_t}{\sqrt{C(H_t)}}
{\rm Re}\Big\{  \int_{\mathbb{R}}
 \frac{e^{- i \xi t}-1}{|\xi|^{1/2+H_t}}  d\tilde{W}(\xi) \Big\} ,
\end{equation}
where the complex random measure $d\tilde{W}$ is of the form $d\tilde{W} = dW_1+idW_2$ with 
$dW_1,dW_2$ two independent real-valued Brownian  
measures, 
and $C(h)$ is the normalization function:
\begin{equation}
C(h) =  
\int_\mathbb{R} \frac{4 \sin^2(\xi/2)}{|\xi|^{1+2h}} d\xi 
= 
\frac{\pi}{h \Gamma(2h) \sin(\pi h)} .
\end{equation}

Let $h\in (0,1)$ and $s \in (0,\infty)$. If $H_t \equiv h$ and $\sigma_t\equiv s$, 
then $B^{(h,s)}(t)\equiv B_{H,\sigma}(t)$ is a fractional Brownian motion with Hurst exponent $h$
and volatility $s$, i.e. 
a zero-mean Gaussian process with covariance
\begin{equation}
\mathbb{E} \big[ B^{(h,s)}(t) B^{(h,s)}(t') \big]
=
\frac{s^2}{2}  
\big( |t|^{2h}+|t'|^{2h} -|t-t'|^{2h}\big).
\end{equation}

Let $\beta \in (0,1)$. Let $H : \mathbb{R} \to (0, 1)$ and $\sigma : \mathbb{R} \to (0,\infty)$ be two $\beta$-H\"older functions,
such that $\sup_t H_t < \beta$.
The multifractional Brownian motion (\ref{def:mbm}) is a zero-mean continuous Gaussian process
that satisfies the Locally Asymptotically Self-Similar property  \cite{benassi97}:
At any time $\tau \in \mathbb{R}$,  we have
\begin{equation}
\lim_{\epsilon \to 0^+} {\cal L} \Big(  \big( \frac{B_{H,\sigma}(\tau+\epsilon t) - B_{H,\sigma}(\tau)}{\epsilon^{H_\tau}}\big)_{t\in\mathbb{R}} \Big)
={\cal L} \Big( 
\big( B^{(H_\tau,\sigma_\tau)}(t) \big)_{t\in \mathbb{R}} \Big) ,
\end{equation}
which means that there is a fractional Brownian motion with Hurst index $H_\tau$ and volatility $\sigma_\tau$
tangent to the multifractional Brownian motion $B_{H,\sigma}$. This implies that its pointwise H\"older regularity is determined by its Hurst exponent.

\section{ Details of Estimation Procedure}\label{sec:appA}

We explain the details of how the scale spectrum and the local power-law parameters are estimated from the price data.

\subsection{Input Parameters}

The  input  parameters are the integers $j_{\rm i}<j_{\rm e}$ that determine the scale range under consideration,
the inertial range, and  the window size $M$  which is the size of the (moving) time window
in   which the local spectra are computed. We must have $1 \leq j_{\rm i} < j_{\rm e} \leq \lfloor M/2 \rfloor$.
 In addition, the  price data  at times  $t_n$  denoted by
$\{ P(t_n), n=1,\ldots,N\}$ are given, where $t_n=t_0+n \Delta t$.

It is possible to estimate the power-law parameters for all center times $t_{n_0}$, $n_0=1,\ldots,N$.
We first give in the next subsection 
the general algorithm for the interior center times $n_0 \in  \lfloor M/2  \rfloor   ,\ldots,N-M+\lfloor M/2 \rfloor$,
and then we extend it to the boundary center times $n_0 \in \{ 1,\ldots,\lfloor M/2 \rfloor  -1\}  \cup \{ N-M+\lfloor M/2 \rfloor+1 ,\ldots,N \} $.

\subsection{Interior Center Times}

Let us fix the {center-time index} 
  $n_0\in  \lfloor M/2 \rfloor+1,\ldots,N-M+\lfloor M/2 \rfloor$ and proceed as follows:

\begin{enumerate}
\item 
Compute the log-transformed data ${\itbf Q} =(Q_j)_{j=1}^M$ centered at $t_{n_0}$ by:
\begin{eqnarray}\label{Y}
Q_j  =  
\log (P(t_{n_0-\lfloor M/2 \rfloor +j} )) , ~~j
=1,\ldots,M .
\end{eqnarray}

\item
Compute the scale spectrum
${\itbf S}=(S_j)_{j=j_{\rm i}}^{j_{\rm e}}$ as the local mean square of the wavelet coefficients:
\begin{eqnarray}\label{eq:scs}
S_j=\frac{ \sum_{i=1}^{N_j} \left(d_j(i)  \right)^2 }{ N_j } ,
\end{eqnarray}
where
\begin{eqnarray}
   N_j         &=&    M - 2j + 1,   \\
   d_j(i)       &=&    \frac{\sum_{k=0}^{j-1} 
   (Q_{k+i} -Q_{k+i + j})}{ \sqrt{2 j}  }  .
   \label{def:dji}
\end{eqnarray}

\item
Define the $(j_{\rm e}-j_{\rm i}+1)$-dimensional vector $\itbf Y$, the $(j_{\rm e}-j_{\rm i}+1)\times 2$-dimensional
matrix ${\bf X}$, and  the $(j_{\rm e}-j_{\rm i}+1) \times ( j_{\rm e}-j_{\rm i}+1)$-dimensional diagonal matrix~$ {\bf R}$
\begin{eqnarray}
\label{def:Y}
{\itbf Y}  &=&  \big( \log_2(S_{j_{\rm i}}),\cdots,\log_2(S_{j_{\rm e}}) \big)^T, \\
 \label{def:designmatrix}
   {\bf X}   &=&
   \left[
  \begin{array}{cc}
                           1 & \log_2(2 j_{\rm i}) \\
                           1 &\log_2(2(j_{\rm i} + 1) ) \\
                           \vdots & \vdots \\
                           1 & \log_2(  2 j_{\rm e} )
                \end{array}
               \right]  ,
\\
                \label{def:hatR}
{R}_{j_1j_2} &=&  {{j_1}} {\bf 1}_{j_1}(j_2) , \quad j_1,j_2\in\{ j_{\rm i},\ldots,j_{\rm e} \}.
\end{eqnarray}

\item
Compute the regression parameters $\hat{\itbf b } = ( \hat{c}, \hat{p})^T$  defined by
\begin{equation}
\label{eq:regb}
\hat{{\itbf b}} = ({\bf X}^T {\bf R}^{-1} {\bf X})^{-1} {\bf X}^T {\bf R}^{-1} {\itbf Y} .
\end{equation}

\item
Compute the local Hurst exponent and volatility estimates as
\begin{eqnarray}
\label{eq:defhatH}
\widehat{H}(t_{n_0})  &=&  \frac{\hat{p}-1}{2}  , \\
\widehat{\sigma}(t_{n_0})  &=&  \frac{2^{\hat{c}/2}}{\sqrt{h(\widehat{H}(t_{n_0}))}}  .
\label{eq:defhatsigma}
\end{eqnarray}
\end{enumerate}
In (\ref{eq:defhatH}) we may threshold the estimation by $\min( \max( \widehat{H}(t_{n_0}) ,0.05) , 0.95)$
in order to avoid any singular behavior.\\
In (\ref{eq:defhatsigma}) the  scale-spectral scaling function $h$ is defined by
\begin{equation}
\label{hfunc}
h(H) =  \frac{(1-2^{-2H})}{(2H+2)(2H+1)} .
\end{equation}
In (\ref{def:dji}) the $d_j(i)$'s correspond to the so-called ``continuous transform Haar
wavelet  detail  coefficients''. \\
In (\ref{def:Y}-\ref{def:hatR}) the matrix ${\bf X}$ is the design matrix, ${\bf R}$ is the least squares weighting matrix, and ${\itbf Y}$ is the data vector for the generalized least squares problem that allows
to identify the local power-law parameters.\\
We remark that the computed volatility is the volatility on the $\Delta t$ time scale.
The local volatility on the time scale $\tau=m\Delta t$ is $
    \widehat{\sigma}(t_{n_0})   m^{\widehat{H}(t_{n_0})} $.

%
%

\subsection{Boundary Center Times}
The spectral information can also be computed for center-time index 
$n_0 \in \{ 1,\ldots,\lfloor M/2 \rfloor -1 \}  \cup \{ N-M+\lfloor M/2 \rfloor+1 ,\ldots,N \} $,
but this requires to narrow the time window used to estimate the local power-law parameters.\\
For  $n_0 \in \{ 1,\ldots,\lfloor M/2 \rfloor -1 \}$, one can apply the algorithm for interior center times
up to the following modifications: First replace $M$ by
$\tilde{M} = n_0- \lfloor M/2\rfloor +M $.
Second replace the data vector ${\itbf Q}$ in Eq. (\ref{Y}) by $\tilde{\itbf Q} = (\tilde{Q}_j)_{j=1}^{\tilde{M}}$ defined by
\begin{eqnarray}
\label{Y2}
\tilde{Q}_j  =  
\log (P(t_{j})) , ~~j
=1,\ldots,\tilde{M} .
\end{eqnarray}
The center times 
$n_0 \in  \{ N-M+\lfloor M/2 \rfloor+1 ,\ldots,N \} $
can be treated similarly. 

\subsection{Summary}
We make explicit the dependence on the inertial  range and time window $j_{\rm i}, j_{\rm e}, M$ and the result of the above
procedure is then: \\
 For a given time  series of prices 
$\{P(t_n), n=1,\ldots,N \}$, for each center time $t_{n_0}$,  $n_0 \in \{ 1,\ldots,N\}$,
we obtain  the associated scale spectrum, the  local Hurst exponent, and the local volatility:
 \begin{eqnarray}
   && S(j,t_{n_0}; M), \quad  j \in \{j_{\rm i},\ldots, j_{\rm e}\}, \\
&& \widehat{H}(t_{n_0};j_{\rm i}, j_{\rm e},M),\quad 
\widehat{\sigma}(t_{n_0};j_{\rm i}, j_{\rm e},M)   .
\end{eqnarray}

\subsection{Remarks on Time Window and Inertial Range Parameters}

The time window size $M$ is chosen according to the scale of information one wants to probe.
In Figure \ref{fig:3} it was chosen to include the full time series in which case one obtains  the 
``global'' scale spectral model. In Figure \ref{fig:6} it was chosen as $M=2^8$ to resolve 
the local variations in Hurst exponent and volatility.  Similarly the inertial range $[j_{\rm i}, j_{\rm e}]$ is
chosen according to which scales  we look for a power  law. 
\re{ In the examples shown in this  paper
we used $j_{\rm i}=1$ and $j_{\rm e}= \lfloor  M/2 \rfloor  $. This is the maximum inertial range given the window size.  }
If the data exhibit  a power law only in  a limited inertial range then indeed 
 $j_{\rm i}, j_{\rm e}$ would be chosen accordingly. 

We also remark that the inertial range and time window size could have been chosen via  an automatized 
procedure. For  instance by choosing several window sizes to get an a priori estimate
of local scale of variation of the parameters, $\sigma,H$  and then subsequently choosing a local window size
to maximize a signal-to-noise ratio.  Similarly, the inertial range, given a time window size $M$, could be chosen
as the maximal inertial range for which the fitting residual conforms with that associated with a  Gaussian power law model.

\subsection{Remarks on Optimality,  Precision and Robustness} 
 
\begin{figure}
   \centering
    \includegraphics[width=6.0cm]{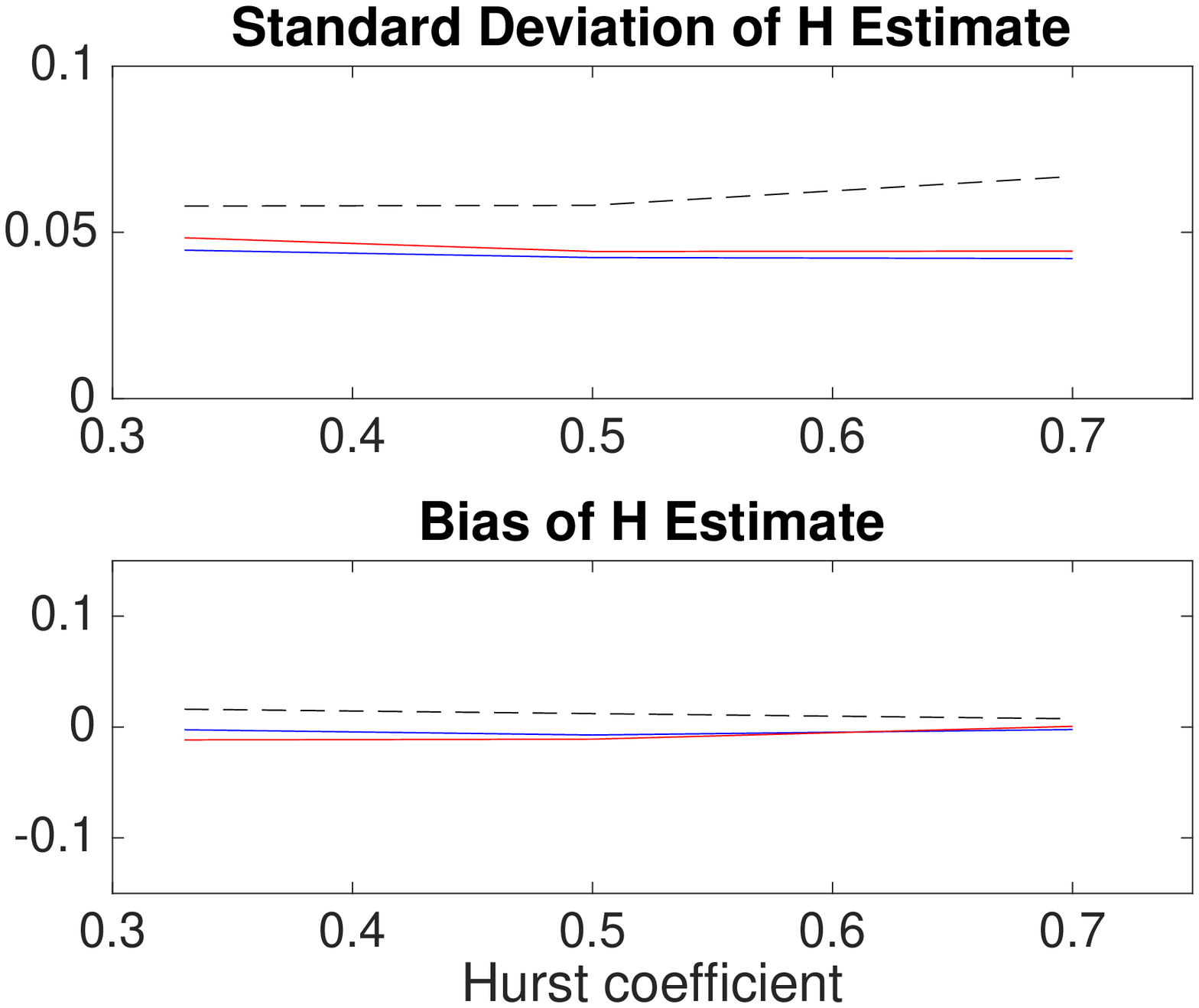}   
   \includegraphics[width=6.0cm]{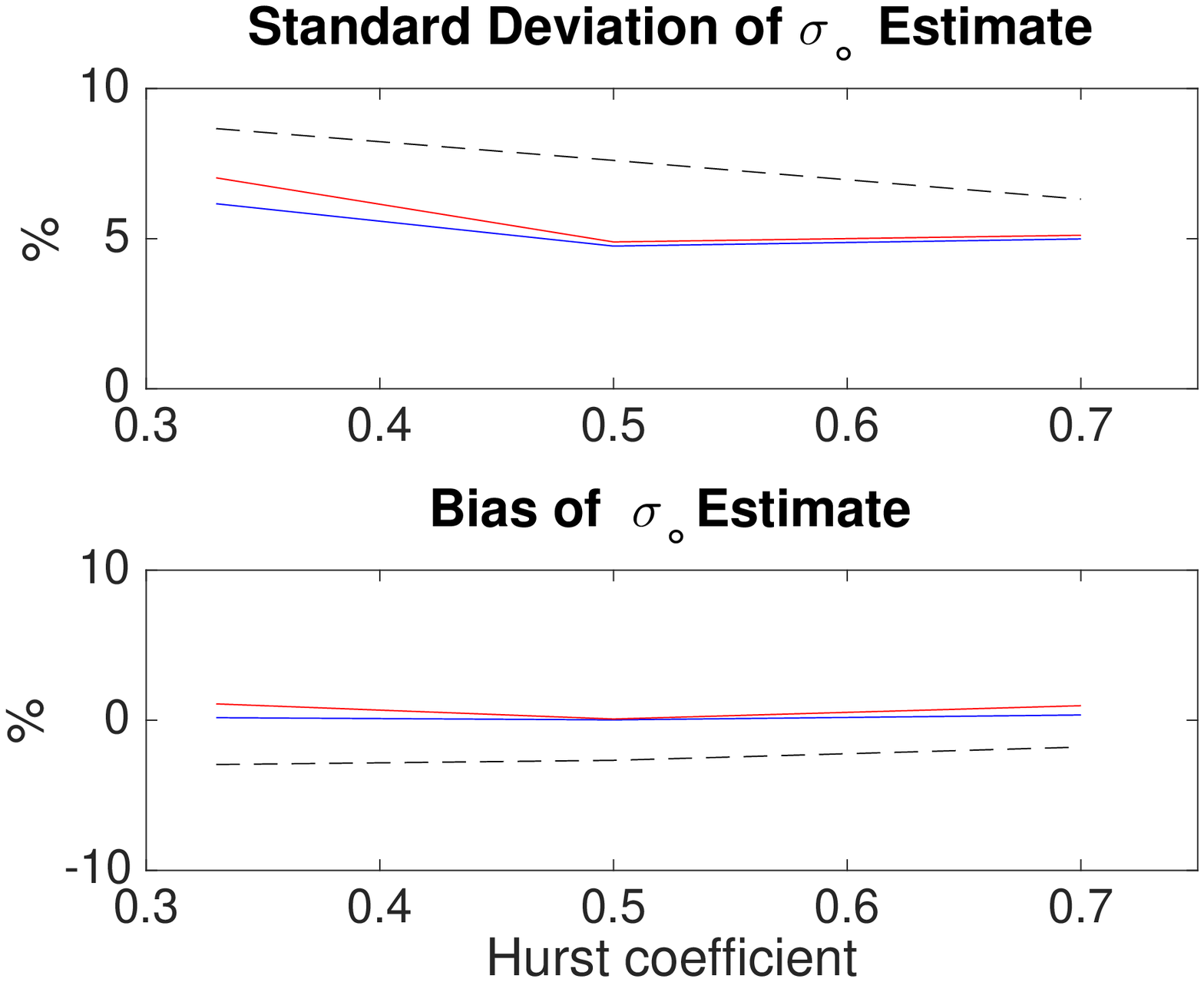}
     \caption{Performance of ML estimator (solid blue line), scale-based estimator
 with full covariance matrix (solid red line), and robust estimator (dashed black line) 
 in case of pure fractional Brownian motion.}
   \label{fig:Bprec0}
\end{figure}

\begin{figure}
   \centering
   \includegraphics[width=6.0cm]{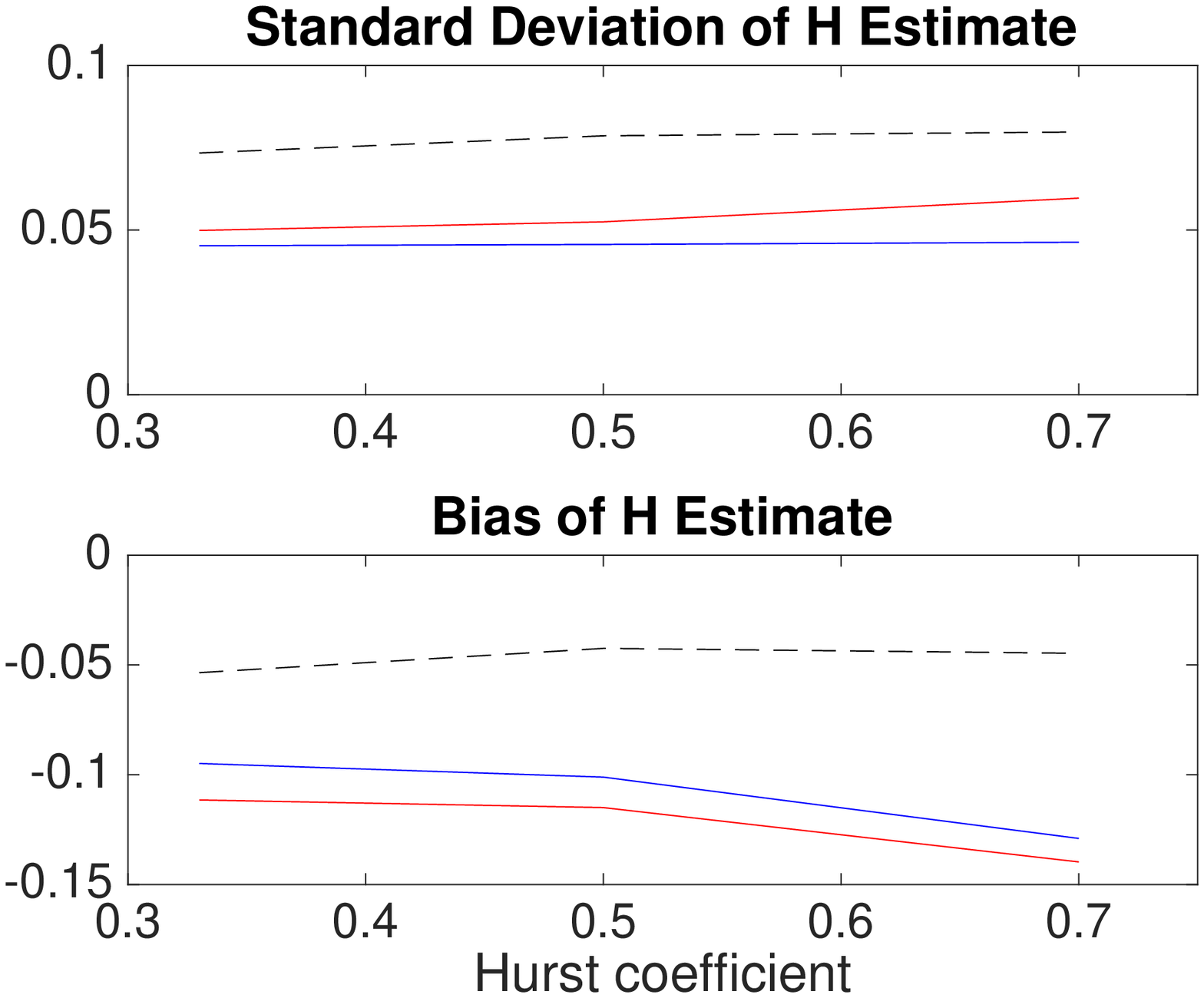} 
   \includegraphics[width=6.0cm]{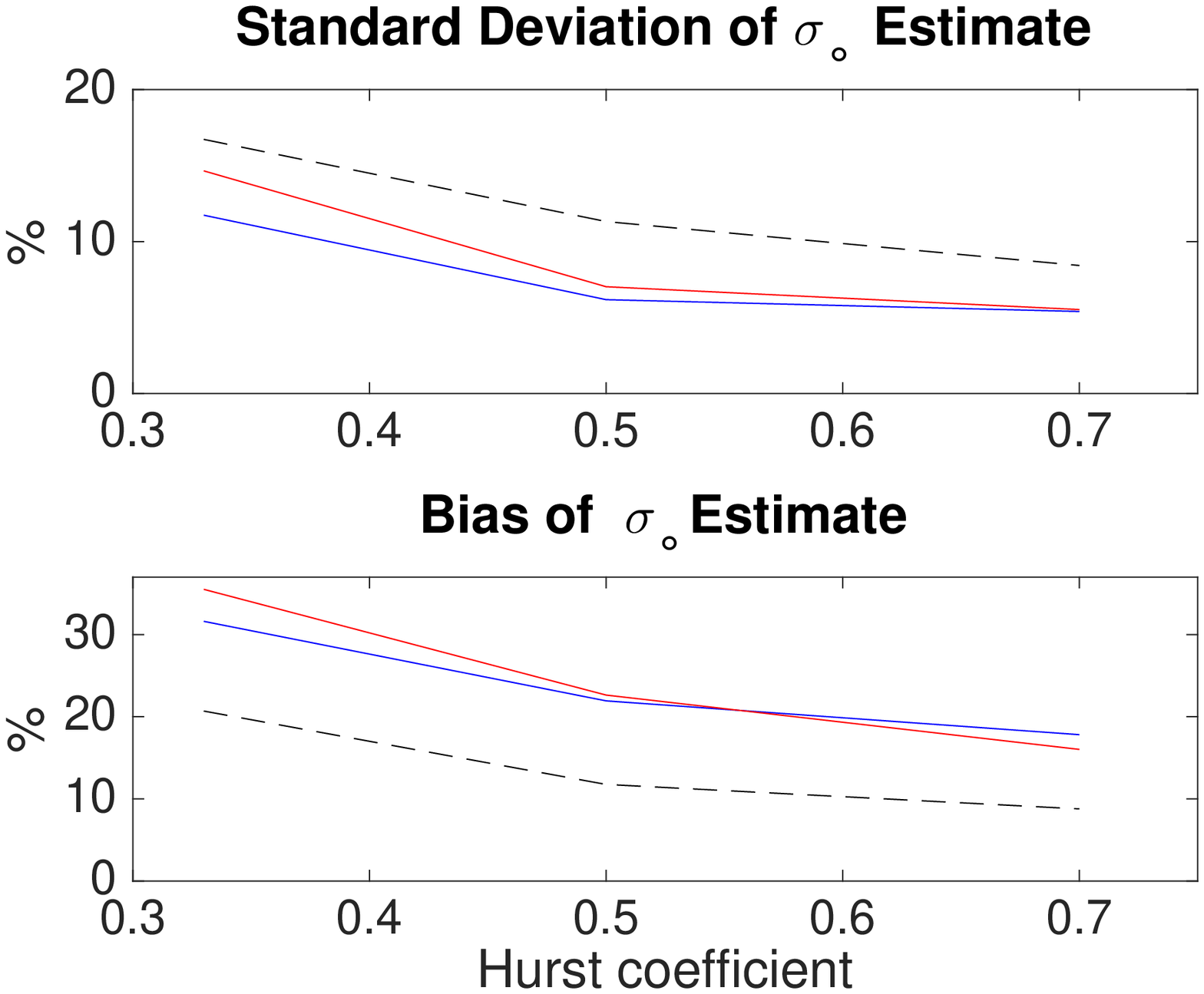}
   \caption{Performance of the estimators 
   in case of fractional Brownian motion  corrupted by white noise.}
   \label{fig:Bprec}
\end{figure}

When the underlying data are pure fractional  
Brownian motion (i.e. a Gaussian process with mean zero and covariance ${\rm Cov}(  B_{\sigma,H}(t),B_{\sigma,H}(s))=
\frac{\sigma^2}{2} (|t|^{2H}+|s|^{2H}-|t-s|^{2H})$) the increments have the form of a Gaussian vector
with correlation matrix determined by the parameters $\sigma$
and $H$. The maximum likelihood (ML) estimator
has been shown to be optimal for high-frequency  point observations of fractional Brownian motion \cite{brouste}.
We show below  that the scale spectral estimator
has essentially the same accuracy as the ML
when the observations come from pure fractional Brownian motion.  
However,  robustness 
in the context of ``imperfect'' data and multiscale estimation are important issues in our context,   
that  is why we use the   estimator  
described above with a diagonal covariance matrix (\ref{def:hatR}) used in the linear 
regression. Indeed we show below that it is much more robust with synthetic data corrupted 
with additive white noise or with real data.
 We remark that the diagonal  weight matrix 
 ${\bf R}$ in the least squares regression in 
 Eq. (\ref{eq:regb})  is important.  
 In order to obtain a robust estimator we use  an estimate that
 derives from two different diagonal scalings. To make the
 dependence on the diagonal weight matrix explicit we denote
 the estimator in Eq.~(\ref{eq:defhatH}) by $\hat{H}(t_{n_0},{\bf R})$.
 Then we  choose the actual estimator as:
 \begin{eqnarray}
 \hat{H}(t_{n_0})  = \max\{ \hat{H}(t_{n_0},{\bf R}), \hat{H}(t_{n_0},{\bf R}^{(3)}) \}    ,
  \end{eqnarray}
 with  ${\bf R}$ defined by Eq. (\ref{def:hatR}) and  ${\bf R}^{(3)}$ defined by
  \begin{eqnarray}
  {R}^{(3)}_{j_1j_2} &=&  {{j_1^3}} {\bf 1}_{j_1}(j_2) , \quad j_1,j_2\in\{ j_{\rm i},\ldots,j_{\rm e} \}.
 \end{eqnarray}
 We illustrate the robustness of this approach  next. 
In Figure \ref{fig:Bprec0} we show the performance of the 
estimation schemes in the case
of synthetic data that have the distribution of a fractional  
Brownian  motion. 
\re{The length of the data vector used here is the one used to  
 compute the local power  spectra for 
  the oil price and gas data and is $2^8=256$ corresponding
to a dyadic dimension of $8$ and to a one year window size approximately.}
The dashed  black line is the robust estimator with diagonal covariance matrix, the red line is the scale-based estimator
 with full  covariance matrix, and the blue line  is the  ML estimator.  
 It clearly appears that the ML estimator is more efficient, both in terms of  biases and variances.
 In Figure \ref{fig:Bprec} we show the performance
when the same data are corrupted by additive measurement noise. 
It can be seen that the ML estimator and the scale spectral
estimator using the full covariance matrix have strong biases particularly
 for high values of the Hurst exponent $H$. Such estimators would under-estimate such high values
 of $H$ when applied to real data.
 In contrast the robust estimator has a much lower bias and should be used with real data.
    
\begin{figure}
   \centering
   \includegraphics[width=6.0cm]{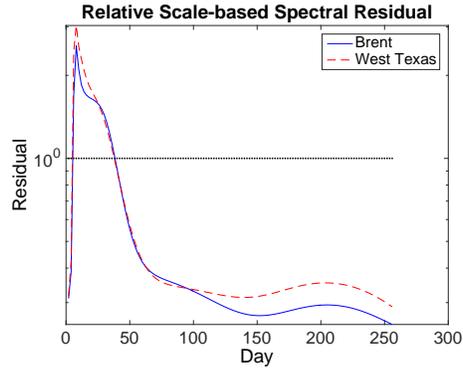} 
   \caption{Relative spectral residual (ratio of the residual obtained with the robust estimator 
   and the one obtained with the ML estimator) 
    as a function of scale for the Brent (solid blue line) and West Texas (dashed red line) data sets. 
}
   \label{fig:Brelres}
\end{figure}

In  Figure \ref{fig:Brelres} we show  the relative spectral residuals
as a function of scale for the Brent and West Texas data sets. 
The relative residual is the ratio of the mean square difference between the empirical spectrum 
and the spectrum estimated with the  robust estimator and the mean square difference between the empirical spectrum 
and the spectrum estimated with the ML estimator. 
  The robust scheme shows superiority at the very short as well as at the
 long scales with a ``spectral cross-over'' at about the scale of a month,
 showing a more consistent tracking of the spectral behavior by the robust scheme.
 We seek here a  robust spectral estimate  for a  
wide-range of scales  in particular  for the
 longer scales that may be important in applications and the figure shows that
 for the long scales  the robust estimator capture much better
 the spectral character.   
 
\begin{figure}
   \centering
    \includegraphics[width=6.0cm]{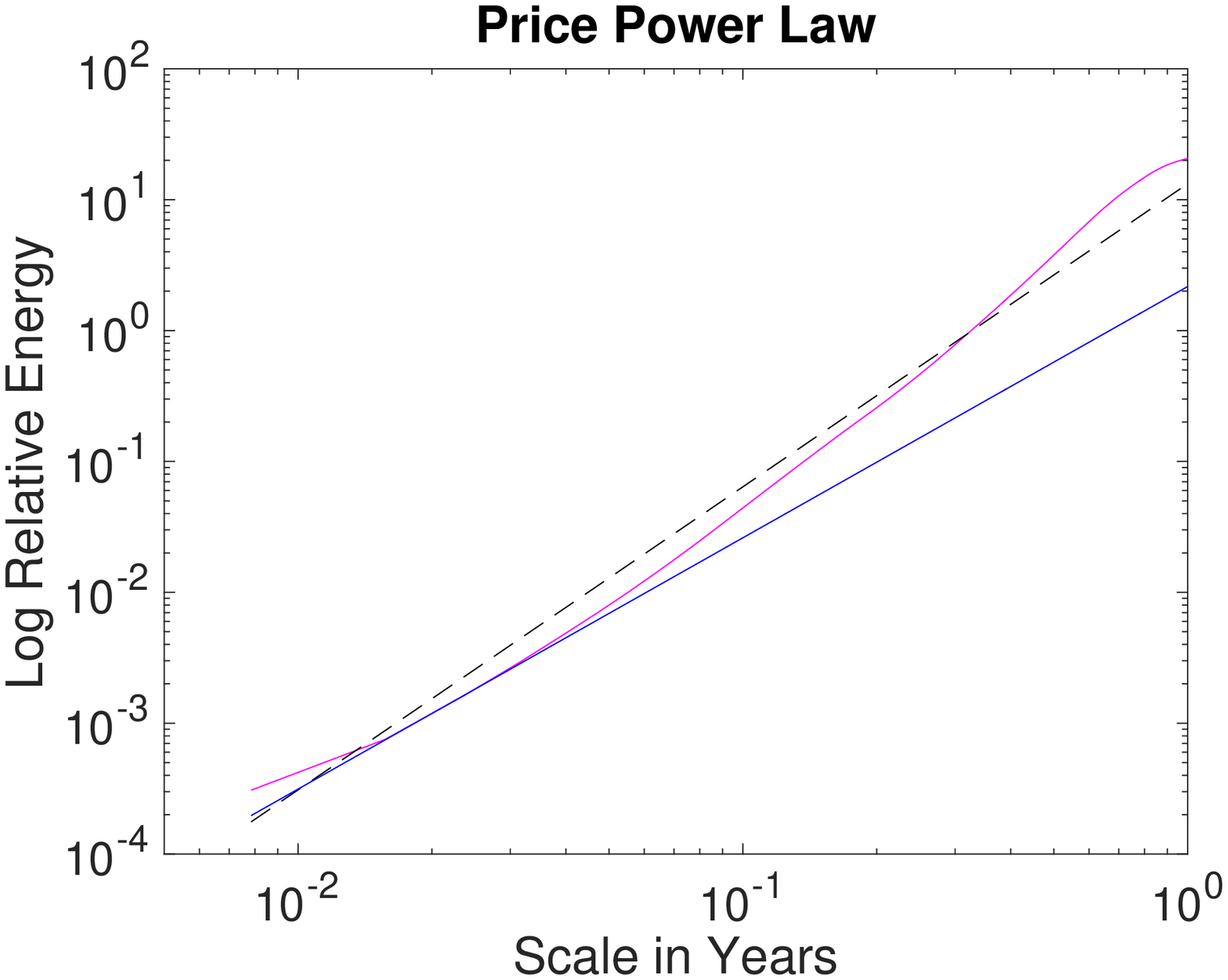} 
    \includegraphics[width=6.0cm]{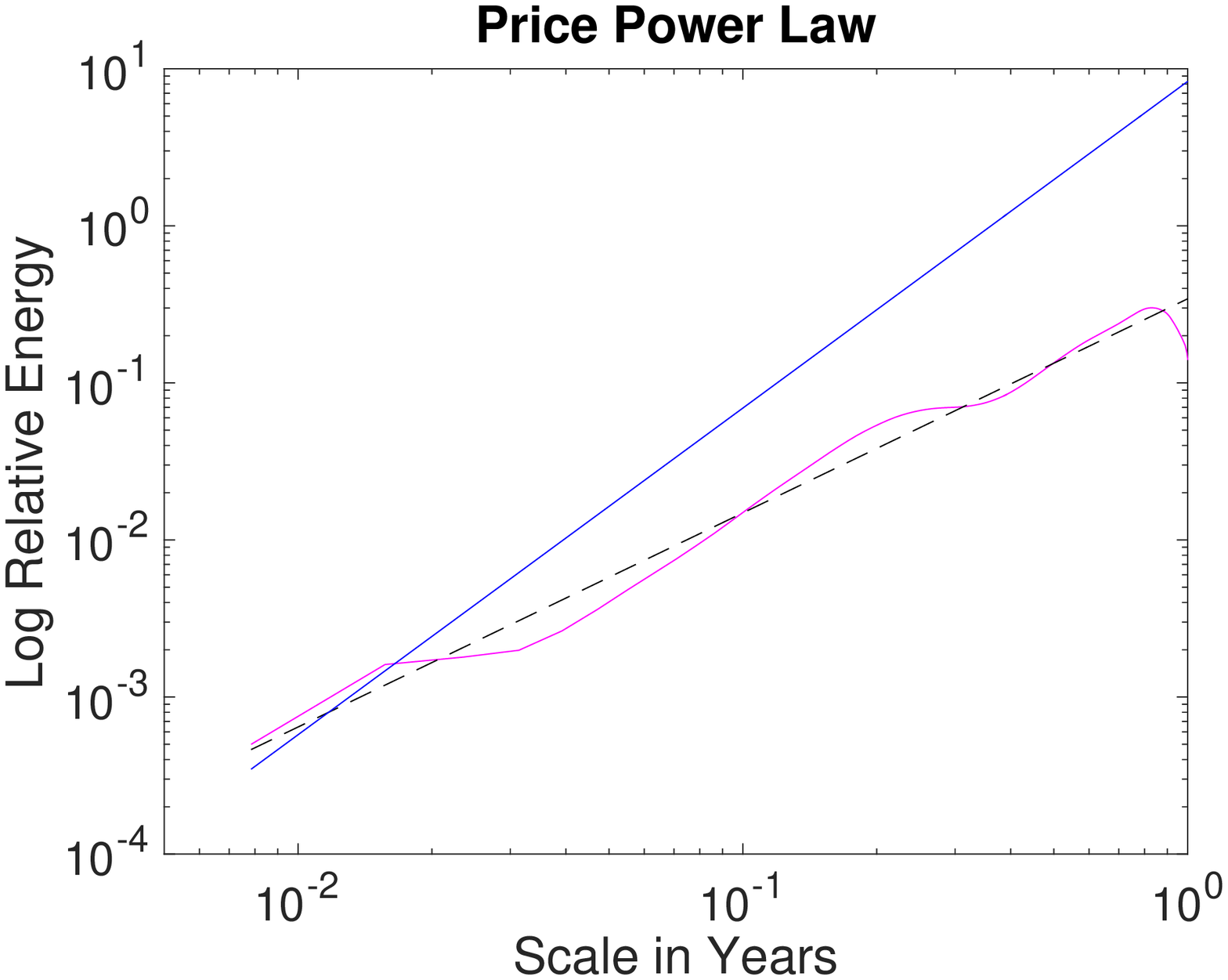}
   \caption{Examples of scale spectral fit for 
   the robust (straight dashed black line) and the ML (straight solid blue line) estimation schemes, compared to the actual spectra (solid red line).
   The left  plot is a local spectrum obtained from the West Texas data and the right  plot is a local spectrum obtained from the Brent data.
   }
   \label{fig:Bspec}
\end{figure}

In Figure \ref{fig:Bspec} we show the actual local scale spectra  
by solid lines, the robust  estimates by the dashed  lines, and the ML  estimates  
by the straight solid lines 
for two cases where the Hurst exponent estimate is high, respectively low.
In the first case (top figure, West Texas) the estimates are $H=.7, \sigma=82\%$
for the robust estimator and $H=.5, \sigma=32\%$ for ML estimator. 
In the second case (bottom figure, Brent) the estimates are $H=.2, \sigma=14\%$
for the robust estimator and $H=.5, \sigma=63\%$ for ML estimator. 
We can explain the dramatic differences between the two estimates by the fact that
the ML estimator is sensitive to small scales and to high-frequency noise. If the 
underlying process is a pure fractional Brownian motion, then the ML estimator performs very well, 
but if the spectrum of the process is only approximately a power law for instance, 
then the ML estimator may be biased.

\section{Further Oil Price Figures}
\label{sec:appB}
 
  \begin{figure}
   \centering
   \includegraphics[width=6.0cm]{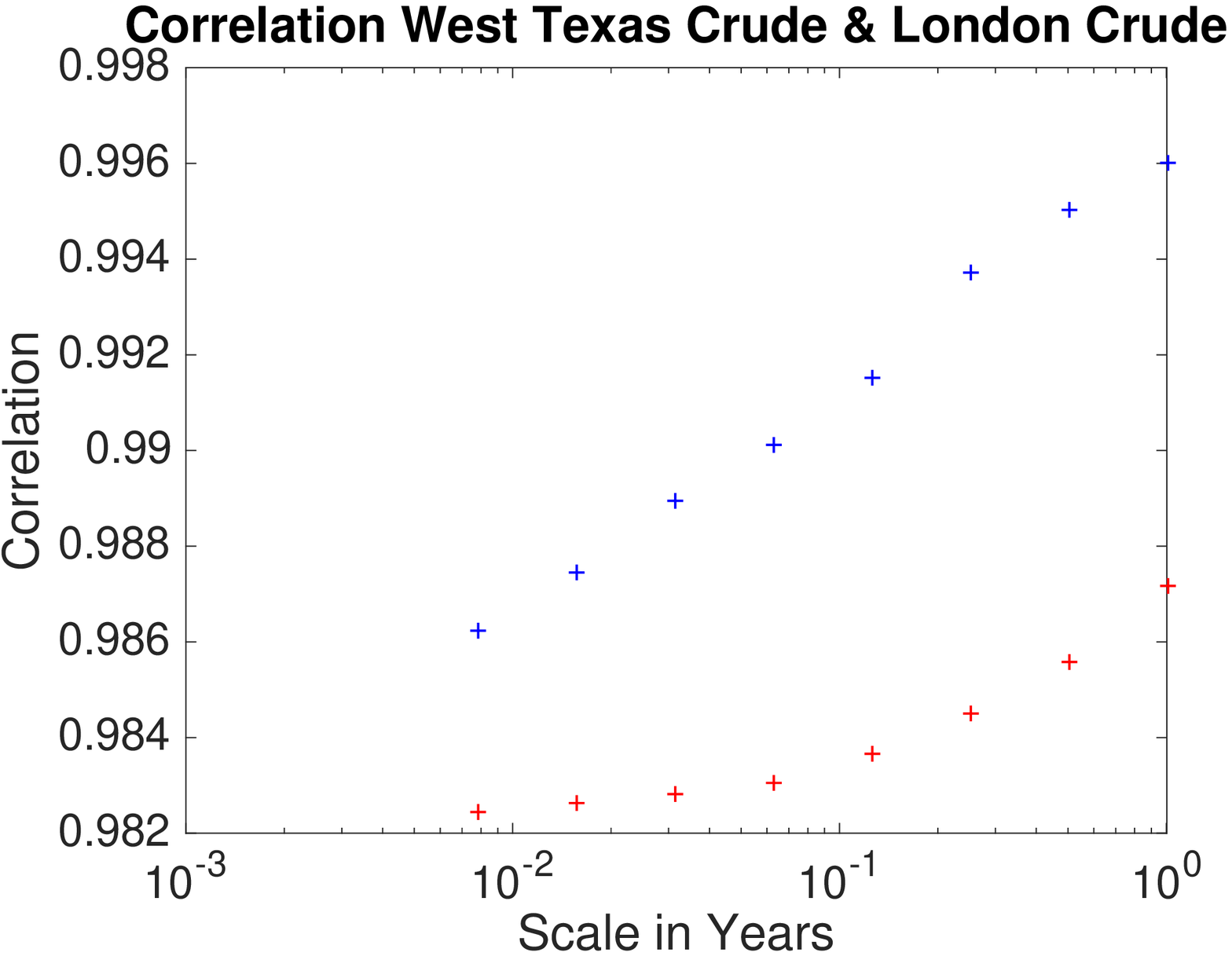}    
   \caption{Scale-based correlation between the Brent and West Texas data
      during the period May 1987--June 2009 (blue crosses) and July 2009-June 2016 (red crosses). }
   \label{fig:corr}
\end{figure}

\begin{figure}
   \centering
      \includegraphics[width=6.0cm]{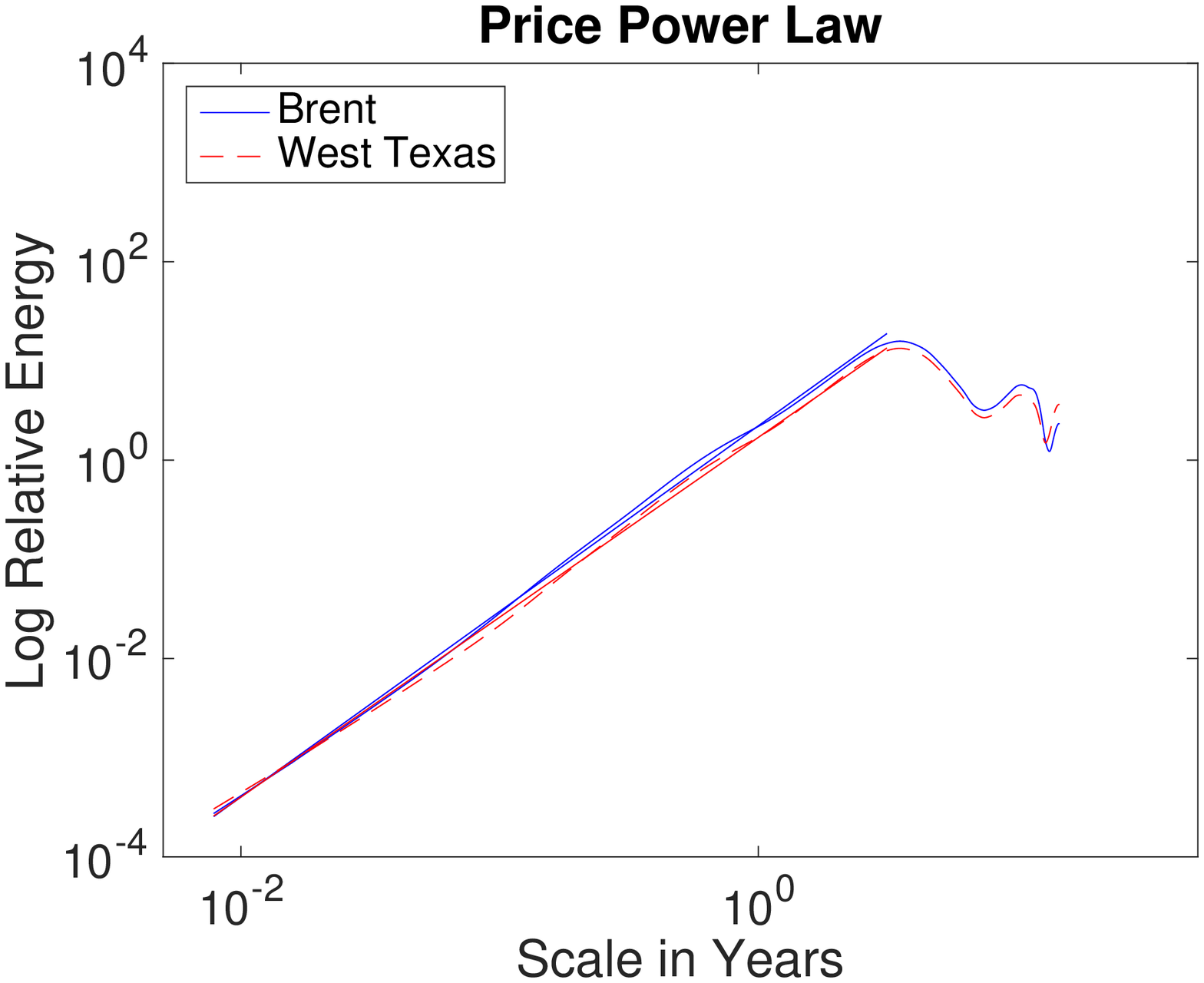} 
            \includegraphics[width=6.0cm]{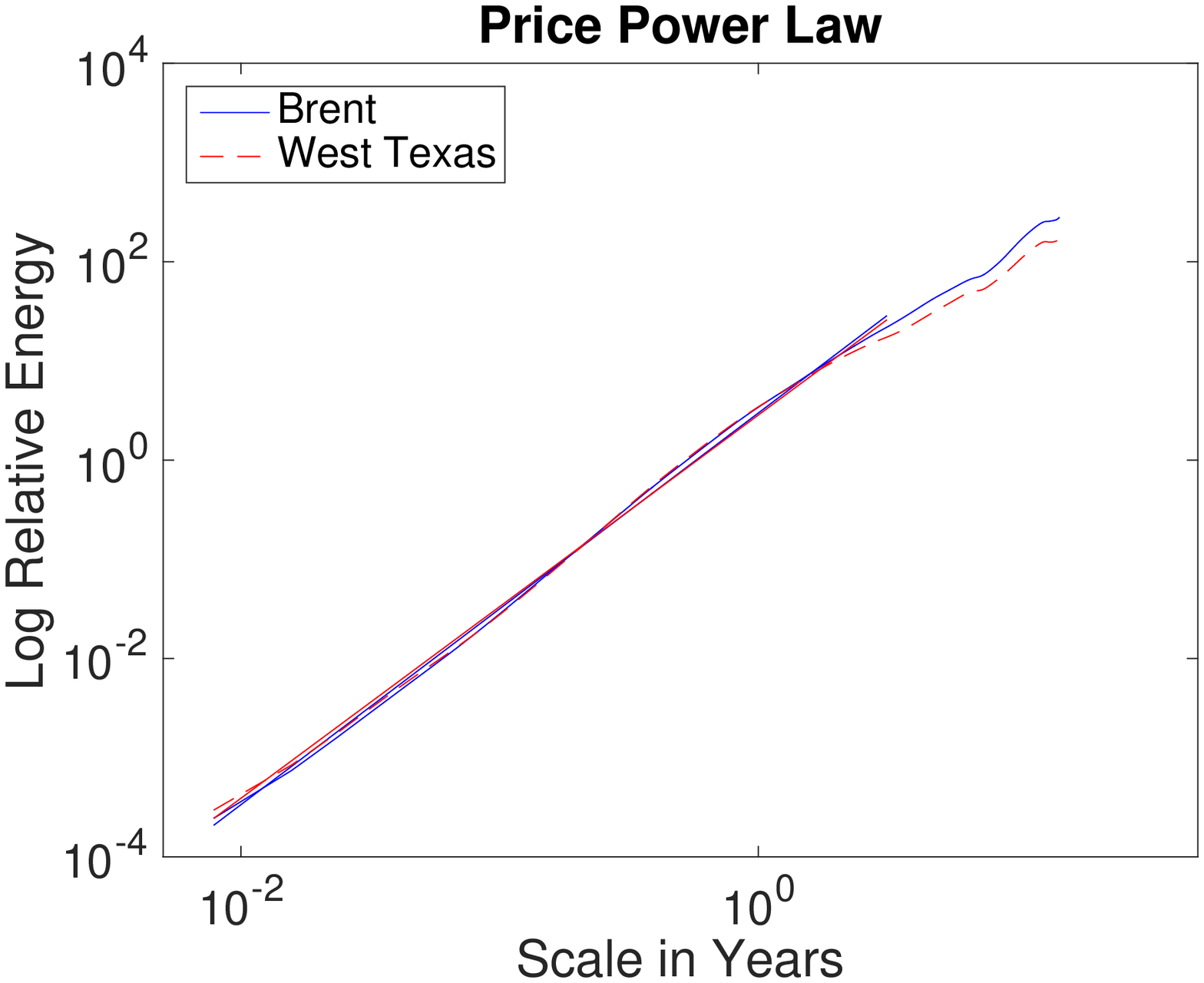}
   \caption{Scale spectra for the period 1987--2002 (left) and  2001--2016 (right).
For the period 1987--2002  the estimated Hurst exponents are  $H=.41$ (West Texas, dashed red line) and $H=.44$ (Brent, solid blue line),
and  the estimated volatilities are $\sigma= 28 \%$ (West Texas) and $\sigma=32 \%$ (Brent).  
The straight  lines are the fitted spectra and fit  well the data up to an outer scale of more than a year.
For the period 2001--2016 the estimated Hurst exponents are $H=.42$ (West Texas) and $H=.45$ (Brent),
and the estimated volatilities are $\sigma=32 \%$ (West Texas) and $\sigma=34 \%$ (Brent).   
 } 
    \label{fig:4}
\end{figure}

\re{
In Figure~\ref{fig:3} we show the scale spectra for the full period of the data while    
 in Figure~\ref{fig:4},  the spectra  correspond to the 
   first and last 16 years of the price data (the periods 1987--2002 and 2001--2016).
   The strength of the price fluctuations is slightly larger  in the latter half of the data for the longer scales.   }
The qualitative behavior of the spectrum  is, however, similar for  the two  halves of the data.
     The Hurst exponent associated with the two halves  are $H=.41$ for the  first 16 years and 
   $H=.42$ for the last 16 years for the West Texas data  (dashed red lines).
    The solid blue lines are the  spectra for the Brent data set with the associated Hurst exponent 
    estimates being $H=.44$ and $H=.45$, respectively.  
    The estimated volatilities are $\sigma=28 \%$ for the  first 16 years and 
   $\sigma=32 \%$ for the last 16 years for the West Texas data  (dashed red lines).
    The estimated volatilities are $\sigma=32 \%$ for the  first 16 years and 
   $\sigma=34 \%$ for the last 16 years for the Brent data  (solid blue lines).
     The straight  lines in the figure are the corresponding model
    spectra that represent a ``perfect'' power law with the estimated exponents.   
   We can observe  that the spectra retain an approximate power-law behavior on very long scales. 
    Moreover, the power law is very similar in the two halves of the data.
     This happens when we estimate the scale spectra over long time periods of 16 years.
      We show in Section \ref{sec:H} that if we consider the data over shorter subwindows,  
     the scale spectra change over time, as we can observe vertical shifts
     in the spectra, corresponding to changes in the local volatility.
     These changes happen in a coordinated fashion for the West Texas and Brent data sets.
      
%
    
\section{Further Natural Gas Price Figures}\label{sec:appC}
In this section we study further the natural gas spot price and
display the corresponding  figures as for the crude oil price data.

\begin{figure}
   \centering
   \includegraphics[width=6.0cm]{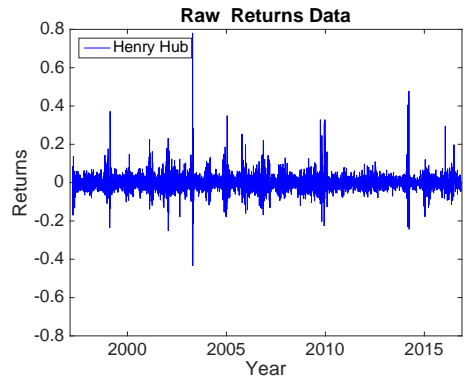} 
   \caption{Returns for Henry Hub natural gas.}
   \label{fig:gret}
\end{figure}

In Figure \ref{fig:gret} we plot the returns, defined as in Equation (\ref{eq:ret}),   
for the Henry Hub natural gas data shown in Figure \ref{fig:gasprice}.
Again the volatility  of the returns process is not constant, but rather exhibits 
temporal variations, moreover, spikes can be seen for special market events.

\begin{figure}
   \centering
      \includegraphics[width=6.0cm]{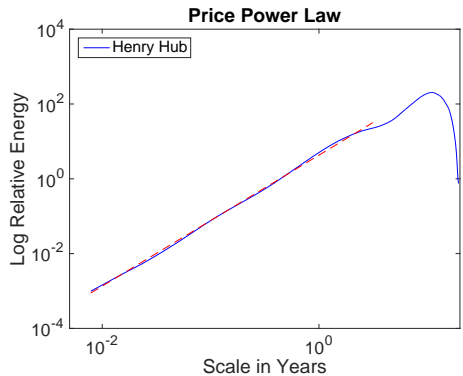} 
   \caption{The  ``global power law'' for Henry Hub natural gas. 
    When the scale spectrum is computed from the complete log-transformed data, we observe
      approximately a power law. The estimated Hurst exponent is $H=.38$.
      The estimated volatility is $\sigma=45 \%$.
     The dashed red  line is the fitted spectrum  and fits
well the data up to an outer scale of more than  one year. 
   }\label{fig:specgas1}
\end{figure}

\begin{figure}
   \centering
      \includegraphics[width=6.0cm]{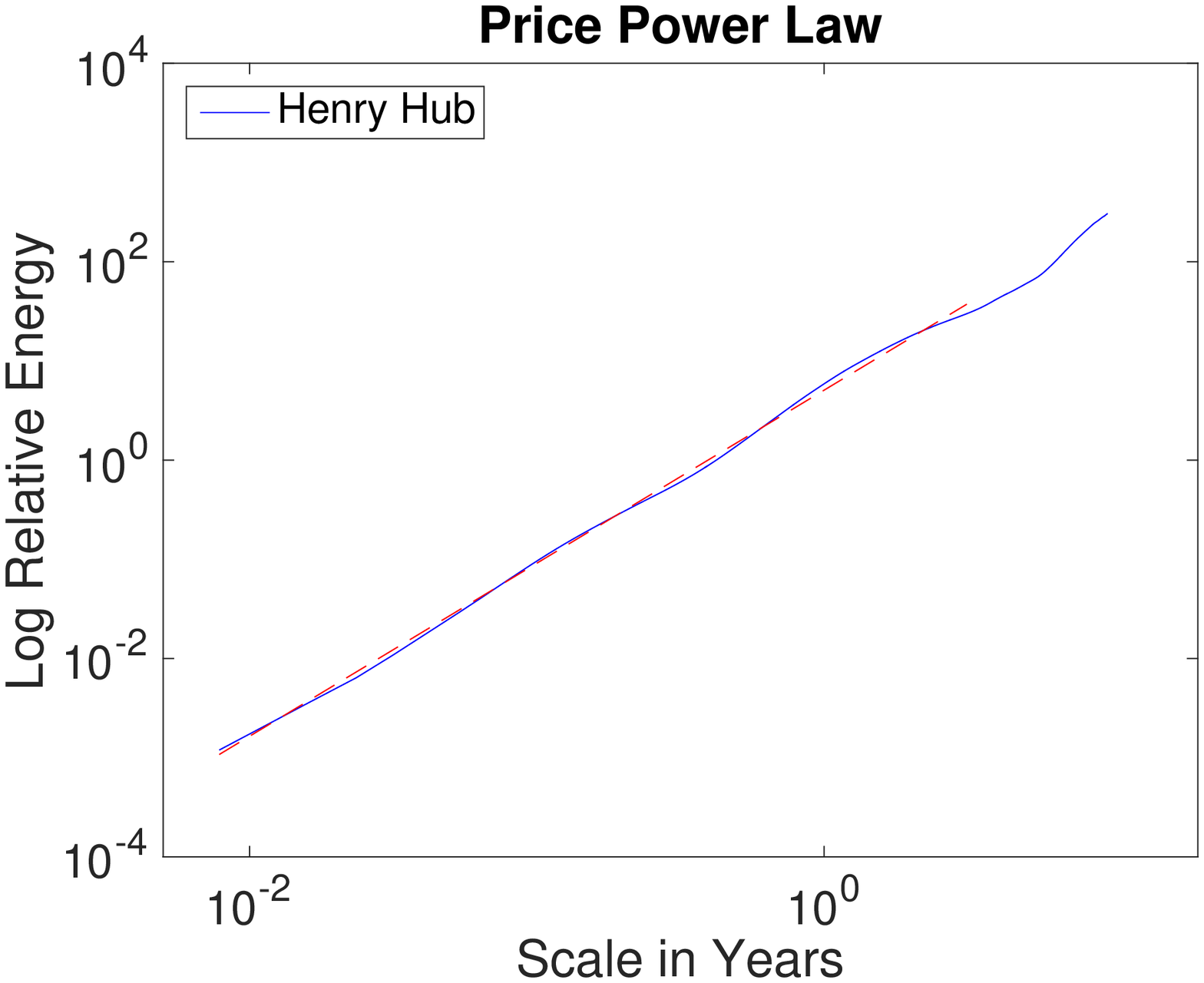} 
      \includegraphics[width=6.0cm]{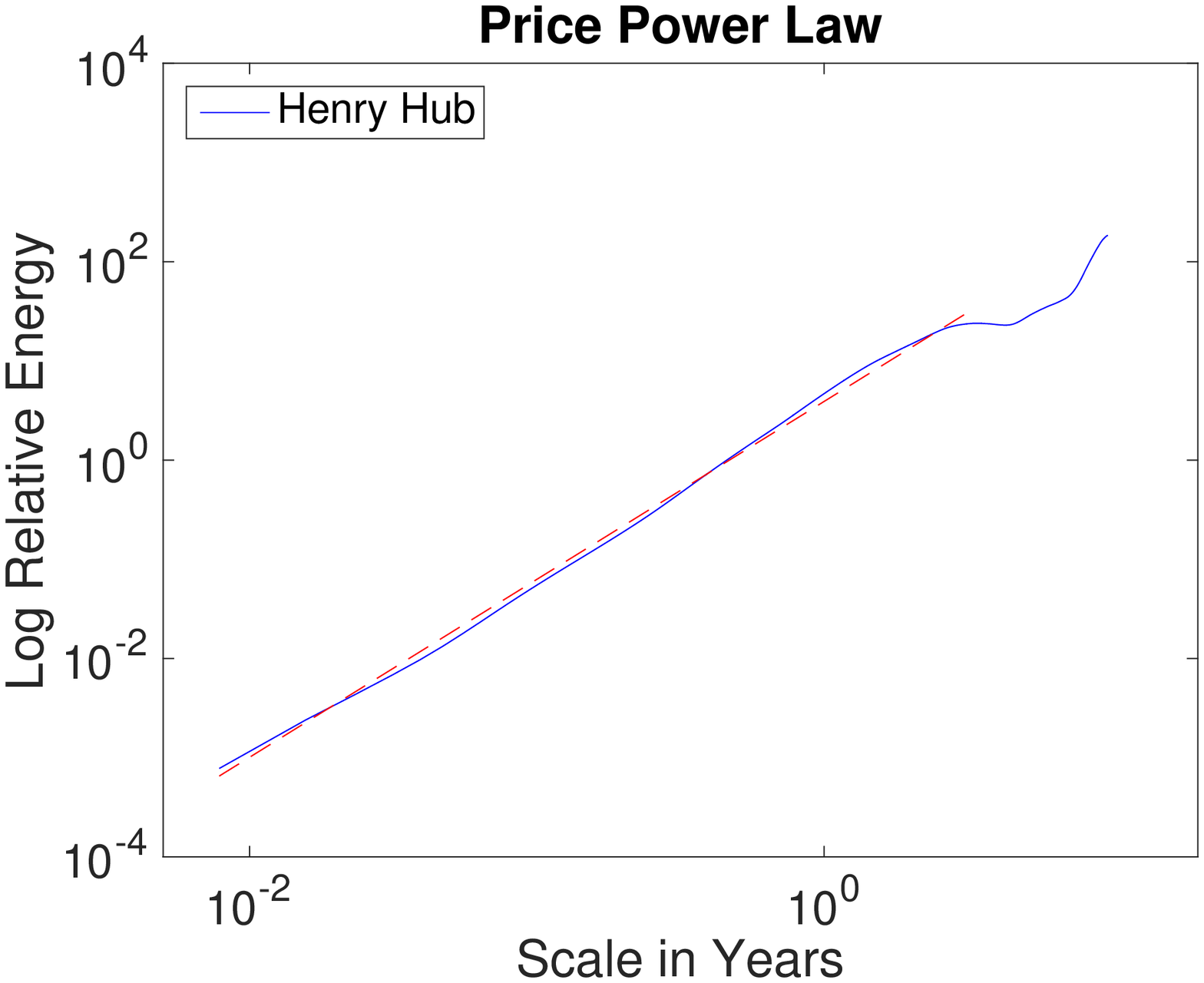} 
   \caption{Scale spectrum for Henry Hub natural gas for the period 1997--2009 (top) and for the period   2009--2016 (bottom).
The estimated Hurst exponent is $H=.37$ (top) and $H=.40$ (bottom).
The estimated volatility is $\sigma= 49\%$ (top) and $\sigma=43 \%$ (bottom).  
The red dotted  lines are the fitted spectra  and fit
well the data up to an outer scale of more than  one year.  } \label{fig:specgas2}
\end{figure}

 Figure \ref{fig:specgas1}  shows  the scale spectrum for the full gas data
 time series. The estimated Hurst exponent is $H=.38$  and 
 the associated   volatility estimate  is  $\sigma=45 \%$.  
 The dashed red  line shows the fitted scale spectrum.
 Note again the linear behavior corresponding to an effective ``mono-fractional'' 
 behavior up to an outer scale of more than a year. The volatility is seen
 to be somewhat higher than for the crude price  fluctuations
 while the Hurst exponent,  and thus the price roughness,  is similar.  
We display corresponding scale spectra in Figure  \ref{fig:specgas2}
 for respectively the first and last half of the data
 (the periods 1997--2007 and 2007--2016).
 The estimated  Hurst exponents are $H=.37$  and $H=.40$ for the first and the last
 halves of the data, while
 the associated volatility estimates  are $\sigma=49 \%$   and 
   $\sigma=43 \%$ respectively.
       The qualitative behavior of the spectrum  is similar for  the two  halves of the data.   
 
We have  calculated the spectral misfit for the natural gas data, that is
root mean square of the difference of the empirical and estimated log-scale spectra
as shown in Figure \ref{fig:7e} in the case of the crude oil data.
As  in  the case of the crude prices we observe  then that the spectral misfit
 is  low and statistically homogeneous with respect to time, which shows that
the time series is   well-described
by the multi-fractional  model with Hurst exponent $H_t$ and $\sigma_t$.
We have moreover considered the situation when we condition on $H=1/2$ 
corresponding to modeling with uncorrelated returns.  The estimated 
volatility path is then more erratic than in the multi-fractional case and
the spectral residual is much larger and with stronger time coherence.  
Thus, again we see that 
 the spectrum is poorly modeled by fixing $H$ as compared
to the multi-fractional case.

 \end{document}